\documentclass[floatfix, aps, prd, superscriptaddress, preprint, preprintnumbers]{revtex4-1}

\usepackage[colorlinks = true, citecolor = blue, urlcolor = blue, linkcolor = blue]{hyperref}
\usepackage[dvipsnames]{xcolor}
\usepackage{pifont} 
\usepackage{amsmath, amssymb}
\usepackage[capitalise]{cleveref} 
\usepackage[e]{esvect} 
\usepackage{esint} 
\usepackage{slashed}
\usepackage{physics} 
\usepackage{enumitem}
\usepackage{graphicx}
\usepackage{tikz}
\usetikzlibrary{arrows.meta} 
\usetikzlibrary[shapes.arrows] 
\usetikzlibrary[shapes.misc] 
\usetikzlibrary{positioning}
\usetikzlibrary{quotes} 
\usepackage[normalem]{ulem}

\usepackage{array, multirow}
\usepackage{bm, bbm}

\newcommand{\xb}{x_{\!_B}}
\newcommand{\xf}{x_{\!_F}}

\newcommand{\fref}[1]{Fig.~\ref{f.#1}}

\newcommand{\eref}[1]{Eq.~(\ref{e.#1})}

\begin{document}
\preprint{JLAB-THY-22-3462}

\title{How well do we know the gluon polarization in the proton?}

\author{Y. Zhou}
\affiliation{Guangdong Provincial Key Laboratory of Nuclear Science, Institute of Quantum Matter,
South China Normal University, Guangzhou 510006, China}
\affiliation{Guangdong-Hong Kong Joint Laboratory of Quantum Matter,
Southern Nuclear \\ Science Computing Center, 
South China Normal University, Guangzhou 510006, China}
\affiliation{Department of Physics and Astronomy, University of California, Los Angeles, California 90095, USA}
\affiliation{\mbox{Department of Physics, William and Mary, Williamsburg, Virginia 23187, USA}}
\affiliation{Jefferson Lab, Newport News, Virginia 23606, USA \\
         \vspace*{0.2cm}
         {\bf Jefferson Lab Angular Momentum (JAM) Collaboration
         \vspace*{0.2cm}}}
\author{N. Sato}
\affiliation{Jefferson Lab, Newport News, Virginia 23606, USA \\
         \vspace*{0.2cm}
         {\bf Jefferson Lab Angular Momentum (JAM) Collaboration
         \vspace*{0.2cm}}}
\author{W. Melnitchouk}
\affiliation{Jefferson Lab, Newport News, Virginia 23606, USA \\
         \vspace*{0.2cm}
         {\bf Jefferson Lab Angular Momentum (JAM) Collaboration
         \vspace*{0.2cm}}}

\date{\today}

\begin{abstract}
We perform the first simultaneous global QCD analysis of spin-averaged and spin-dependent parton distribution functions (PDFs), including single jet production data from unpolarized and polarized hadron collisions.
We critically assess the impact of SU(3) flavor symmetry and PDF positivity assumptions on the quark and gluon helicity PDFs, and find strong bias from these, particularly on the gluon polarization.
The simultaneous analysis allows for the first time extraction of individual helicity-aligned and antialigned PDFs with a consistent treatment of uncertainties.
\end{abstract}

\maketitle

\section{Introduction}

Ever since the discovery by the European Muon Collaboration that only a small fraction ($\lesssim 10\%-30\%)$ of the proton's spin is derived from quarks~\cite{EuropeanMuon:1987isl}, understanding the decomposition of the proton spin into its fundamental components has challenged the hadron physics community for over 3 decades.
Some initial explanations focused on potentially large cancellations from gluonic contributions via the axial anomaly~\cite{Altarelli:1988nr, Carlitz:1988ab}, or from large negative strange quark polarization, in violation of the Ellis-Jaffe sum rule~\cite{Ellis:1973kp}, or from large higher twist effects that could obscure a simple partonic interpretation~\cite{Close:1993mv}.

Subsequent experiments at CERN, SLAC, DESY and Jefferson Lab involving polarized inclusive or semi-inclusive deep-inelastic lepton-nucleon scattering over a broad range of kinematics, as well as polarized proton-proton collisions at RHIC, have provided a substantive body of data that have largely confirmed the original conclusion of a small total quark polarization~\cite{Anselmino:1994gn, Lampe:1998eu, Aidala:2012mv}.
Moreover, recent global QCD analyses, especially ones which do not impose theoretical constraints from SU(3) flavor symmetry~\cite{Ethier:2017zbq}, have typically favored a fairly small strange quark polarization, consistent with zero, effectively sidelining polarized strangeness from playing a significant role in the proton spin puzzle.

The question of where are the missing pieces of the proton spin has inspired studies of other possible sources, such as gluon helicity or quark and gluon orbital angular momentum~\cite{Diehl:2003ny, Leader:2013jra}.
The latter can be related to moments of generalized parton distributions (GPDs)~\cite{Muller:1994ses, Ji:1996nm, Radyushkin:1996nd}, the determination of which has motivated experimental programs of high-energy exclusive reaction measurements, such as in deeply-virtual Compton scattering.
While progress has been made on the theoretical side with improvements in lattice QCD calculations of GPD moments and Compton form factors~\cite{Alexandrou:2017oeh}, the phenomenological information about parton orbital angular momentum is relatively sparse, with experimental programs still largely in their infancy~\cite{Kumericki:2016ehc}.

In contrast, the extraction of gluon helicity has matured to a somewhat more advanced stage, with jet production data available from polarized $pp$ collisions at RHIC~\cite{Abelev:2006uq, Abelev:2007vt, Adamczyk:2012qj, Adamczyk:2014ozi, Adam:2019aml, STAR:2021mfd, STAR:2021mqa, Adare:2010cc}.
Inclusive jet cross sections offer direct sensitivity to the gluon momentum fraction and helicity distributions, without complications associated with final state hadronization~\cite{Jager:2004jh}.
In a seminal 2014 analysis, de~Florian {\it et al.} (DSSV)~\cite{deFlorian:2014yva} used the RHIC jet production data to extract the first clearly nonzero signal for a polarized gluon distribution in the proton for gluon momentum fractions $x$ between $\approx 0.05$ and $\approx 0.2$.
While the determination of the total gluon helicity in the proton is still subject to large extrapolation uncertainties in the unmeasured small-$x$ region, the establishment of a positive gluon polarization was a major milestone in the developing story of the proton spin decomposition.

In parallel developments, the Jefferson Lab Angular Momentum (JAM) Collaboration has recently pioneered advances in global QCD analysis with simultaneous Bayesian Monte Carlo determination of different types of co-dependent distributions, such as PDFs and fragmentation functions~\cite{Ethier:2017zbq, Sato:2019yez, Moffat:2021dji}, as well as polarized and unpolarized PDFs~\cite{Cocuzza:2021}.
The studies found important correlations between the shapes of the inferred unpolarized~\cite{Sato:2019yez, Moffat:2021dji} and polarized~\cite{Ethier:2017zbq} strange quark distributions and inputs assumed for fragmentation functions, along with theoretical biases imposed on the analysis.

In this paper we use the JAM global QCD analysis framework to study the gluon helicity distribution, and in particular the robustness of the extracted signal in view of theoretical assumptions made in previous global analyses~\cite{deFlorian:2014yva, Nocera:2014gqa}.
Most common of these is the imposition of SU(3) flavor symmetry relating the octet and singlet axial charges, as well as positivity constraints on the $x$ dependence of the distributions~\cite{Candido:2020yat, Collins:2021vke}.
In addition, we simultaneously determine both the spin-averaged and spin-dependent PDFs by fitting to lepton and hadron scattering data, including jet production cross sections from unpolarized and polarized $pp$ collisions (and $p \bar{p}$ collisions for spin-averaged scattering).
The simultaneous analysis allows for the first time extraction of individual helicity-aligned and antialigned PDFs with a consistent treatment of uncertainties.

We perform a careful study of various scenarios employing different theoretical assumptions, and find that indeed the sea quark and gluon helicity distributions can depend strongly on the constraints imposed.
In particular, without restricting PDFs to be positive and assuming SU(3) flavor symmetry for the axial vector charges, existing polarized data allow solutions containing {\it negative} gluon polarization, in addition to the standard positive gluon solutions found in previous analyses, giving equally acceptable descriptions of the data.
Interestingly, a similar double solution was also found earlier by the COMPASS Collaboration in an extraction of spin-dependent PDFs from their proton measurements combined with world inclusive DIS data~\cite{Adolph:2015saz}.
We conclude that further input from higher-precision measurements of existing or possibly new observables over a range of kinematics is needed in order to draw firmer conclusions about gluon polarization and its contribution to the proton spin budget.


The organization of this paper is as follows.
In Sec.~\ref{s.theory} we briefly summarize the theoretical foundations on which this work is based, including a discussion of collinear factorization, Mellin space techniques, and PDF parameterizations.
The data analysis framework is presented in Sec.~\ref{s.analysis}, with details about Bayesian inference, Monte Carlo sampling, and the JAM multi-step strategy.
Also included in Sec.~\ref{s.analysis} is a summary of the inputs used in the analysis, surveying the experimental datasets fitted and the theoretical scenarios explored in this work.
The results of the global QCD fits are presented in \cref{s.results}, where we discuss in detail the shapes of the spin-averaged and spin-dependent PDFs, focusing in particular on the determination of the gluon helicity distribution.
We also infer for the first time the PDFs in the helicity-basis from the combined unpolarized and polarized PDF analysis with a consistent treatment of PDF uncertainties.
Finally, Sec.~\ref{s.conclusion} summarizes the results and discusses the implications of our analysis.
In \cref{s.auc} we provide a brief explanation about the statistical tool employing the area under the curve of the receiver operating characteristic used to visualize the results in Sec.~\ref{s.results}.

\section{Theoretical framework}
\label{s.theory}

In this section we outline the main elements of the theoretical framework on which our analysis is based, with a summary of the essential results from collinear QCD factorization, the use of Mellin space techniques for the scale evolution, and the choice of parameterization for the PDFs.

\subsection{QCD factorization}

For our QCD global analysis of spin-averaged and spin-dependent PDFs we consider data on physical observables available from processes for which QCD factorization theorems exist in leading power collinear factorization.
These include unpolarized and polarized inclusive deep-inelastic scattering (DIS) from protons ($p$) and deuterons ($D$) (and $^3$He for polarized), Drell-Yan lepton-pair production in unpolarized $pp$ and $pD$ scattering, and inclusive jet production in unpolarized $pp$ and $p \bar{p}$ collisions, and in polarized $pp$ collisions.
These are summarized in Table~\ref{tab:processes}, where we also indicate the relevant factorization between the short-distance partonic cross sections $(\Delta) \mathcal{H}$ describing the hard scattering in perturbative QCD and the corresponding nonperturbative PDFs, as well as the kinematic variables involved.
The repeated indices $i, j$ are summed over all parton flavors.

\begin{table}
\begin{center}
\caption{Processes studied in this global QCD analysis of spin-averaged ($f_i$) and spin-dependent ($\Delta f_i$) PDFs, including relevant variables and schematic factorization representation.  Here $N$ represents a proton $p$ or neutron $n$, with $n$ extracted from either deuteron $D$ or $^3$He data.}
\begin{tabular}{ l | c | c }
\hline
~Reaction~ & 
~Variables~ & 
~Factorization~
\\ 
\hline  
~{\bf spin-averaged}~ 
& & 
\\
\qquad $\ell + N\ \to\ \ell' + X$  &
~$\xb$, $Q^2$~          &
~$\mathcal{H}_i^{\mbox{\tiny DIS}} \otimes f_i$~
\\
\qquad \qquad inclusive DIS
& & 
\\
\qquad $p + N\ \to\ \ell^+ + \ell^- + X$ 
& $\xf$, $Q^2$ 
& $\mathcal{H}_{ij}^{\mbox{\tiny DY}} \otimes f_i \otimes f_j$ 
\\
\qquad \qquad Drell-Yan lepton-pair production~~ 
& & 
\\
\qquad $p + p (\bar p)\ \to\ {\rm jet} + X$ 
& $y_{\rm jet}$, $p_{\rm jet}^T$ 
& $\mathcal{H}_{ij}^{\mathrm{jet}} \otimes f_i \otimes f_j$
\\
\qquad \qquad inclusive jet production
& & 
\\
\hline
~{\bf spin-dependent}~ 
& & 
\\
\qquad $\stackrel{\to}{\ell} + \stackrel{\to}{N}\ \to\ \ell' + X$
& $\xb$, $Q^2$
& $\Delta\mathcal{H}_i^{\mbox{\tiny{DIS}}} \otimes \Delta f_i$ 
\\
\qquad \qquad polarized inclusive DIS~~ 
& & 
\\
\qquad $\stackrel{\to}{p} + \stackrel{\to}{p}\ \to\ {\rm jet} + X$
& $y_{\rm jet}$, $p_{\rm jet}^T$ 
& ~~$\Delta\mathcal{H}_{ij}^{\mathrm{jet}} \otimes \Delta f_i \otimes \Delta f_j$~~
\\
\qquad \qquad polarized inclusive jet production~~ 
& &
\\ 
\hline  
\end{tabular}
\label{tab:processes}
\end{center}
\end{table}

For DIS, the cross sections are usually given in terms of the Bjorken scaling variable
    \mbox{$\xb = Q^2/2 P \cdot (\ell-\ell')$},
where $P$ is the four-momenta of the nucleon and $\ell$ and $\ell'$ the four-momenta of the incident and scattered leptons, and the four-momentum transfer squared
    $Q^2 \equiv -(\ell - \ell')^2 \approx 2\, \ell \cdot \ell' > 0$.
For the Drell-Yan process, the cross sections are functions of the Feynman scaling variable, defined in terms of the longitudinal components of the lepton pair,
    $\xf = 2 (\ell^+_L + \ell^-_L)/\sqrt{s}$,
where $s$ is the invariant mass squared of the hadronic collision, and 
    $Q^2 \equiv 2 \ell \cdot \ell'$.
For inclusive jet observables, the variables are the rapidity $y_{\rm jet}$ and transverse momentum of the jet $p_{\rm jet}^T$ in the hadronic center of mass frame.

For polarized observables the short distance cross sections $\Delta {\cal H}$ represent the difference between helicity dependent cross sections, 
    $\Delta {\cal H} = {\cal H}^{++} - {\cal H}^{+-}$.
For polarized inclusive DIS, the labels ``$++$'', ``$+-$'' represent the initial state lepton and hadron for having equal and opposite helicities, while for inclusive jet production these correspond to the the helicity configurations of the incident protons.

In our analysis we use the $\overline{\mathrm{MS}}$ scheme for the renormalization group equations, with the strong coupling $\alpha_s(\mu)$ solved numerically using the beta-function at two loops with $\alpha _s (M_Z) = 0.118$ at the $Z$-boson mass scale $\mu = M_Z$.
The spin-averaged and spin-dependent PDFs are evolved at next-to-leading logarithmic accuracy using the DGLAP evolution equations \cite{Dokshitzer:1977sg, Gribov:1972ri, Altarelli:1977zs} by parameterizing only the light quark and gluon distributions at input scale, $\mu = \mu _0 = 1.27$~GeV.
The heavy quark PDFs are generated perturbatively via the evolution equations, and for the physical observables we use the zero-mass variable flavor scheme with a charm quark mass $m_c = 1.28$~GeV and bottom quark mass $m_b = 4.18$~GeV.
All short-distance partonic cross sections are evaluated at next-to-leading order (NLO) accuracy in perturbative QCD.

For the jet cross sections we utilize the NLO partonic cross sections from J\"{a}ger {\it et al.}~\cite{Jager:2004jh}, along with the corrected expression from Refs.~\cite{Kang:2017mda, Werner:private} based on the small cone approximation with the appropriate settings for the jet reconstruction algorithm.
In practice we use the cone \cite{Blazey:2000qt} and $k_T$-type algorithms \cite{Ellis:1993tq, Cacciari:2008gp} to match with the corresponding experimental datasets.

\subsection{Mellin space techniques}

To perform our numerical analysis we in practice require a fast evaluation of the observables and an efficient procedure to solve the evolution equations for the PDFs.
For the latter, we perform the evolution in Mellin space, which admits simple analytical solutions~\cite{Vogt:2004ns}.
For inclusive DIS, the cross section is a one-dimensional convolution of the short-distance partonic cross sections and PDFs, which can be rendered as a Mellin integral,
    $[A \otimes B](x) = \int_x^1 (\dd{z}\!/z)\, A(z)\, B(x/z)$,
with analytic expressions for the hard functions $\mathcal{H}_i^{\mbox{\tiny{DIS}}}$ and $\Delta\mathcal{H}_i^{\mbox{\tiny{DIS}}}$ in Mellin space available in the literature~\cite{Floratos:1981hs}.

In contrast, the Drell-Yan and inclusive jet production cross sections involve a double convolution of the hard cross section and two nonperturbative functions, which in general cannot be rendered as a true Mellin convolution.
Instead, we utilize the Mellin grid approach developed by Stratmann and Vogelsang~\cite{Stratmann:2001pb}, in which the double convolution is written in the form
\begin{eqnarray}
{\cal H}_{ij} \otimes f_i \otimes f_j
&=& \int^1_{x^{\rm min}_1} \dd{x_1} \int^1_{x^{\rm min}_2} \dd{x_2}\, 
    {\cal H}_{ij}(x_1, x_2)\, f_i(x_1)\, f_j(x_2)
\notag\\
& & \hspace*{-2.8cm}
 =\ \frac{1}{(2\pi i)^2}
    \int \dd{N_1} \int \dd{N_2}\, F_i(N_1)\, F_j(N_2)\,
    \bigg[ 
    \int^1_{x^{\rm min}_1} \dd{x_1} \int^1_{x^{\rm min}_2} \dd{x_2}\,
        {\cal H}_{ij}(x_1, x_2)\, x_1^{-N_1}\, x_2^{-N_2}
    \bigg],
\label{eq:dbleconv}
\end{eqnarray}
where $x_1$ and $x_2$ denote the partonic fractions in hadrons 1 and 2, respectively.
In the second line of Eq.~(\ref{eq:dbleconv}) we have replaced the $x$-space PDFs by their Mellin space representations,
    $f(x) = 1/(2\pi i)\int \dd{N}\, x^{-N}F(N)$.
The factors inside the brackets are independent of the PDFs, and can be calculated for all Mellin moments needed for the double inverse Mellin transform.
They contain all of the kinematic dependence of the process within the hard function ${\cal H}_{ij}$ and the limits for the parton momentum fractions $x^{\rm min}_{1,2}$.
The factors can then be the precalculated and stored as lookup tables when evaluating the Drell-Yan lepton-pair and jet production observables.

\subsection{PDF modeling}
\label{s.pdfs}

For the shape of the spin-averaged and spin-dependent PDFs at the input scale $\mu_0$, we use a generic template function defined by
\begin{equation}
\mathrm{T} \pqty{x; \boldsymbol{a}, n}
= \frac{a_0\, x^{a_1} (1-x)^{a_2} \big(1 + a_3 \sqrt{x} + a_4 x\big)}
       {\int_0^1 \dd{x} x^{n + a_1 - 1} (1-x)^{a_2} 
        \big( 1 + a_3 \sqrt{x} + a_4 x \big)}\, ,
\label{e.template}
\end{equation}
where $\boldsymbol{a} = \{ a_0, \ldots, a_4 \}$ is the set of free parameters for each PDF flavor.
The template function is normalized with respect to the $n$-th moment in order to numerically decorrelate the overall normalization parameter $a_0$ from the shape parameters $a_1, \ldots, a_4$.
The spin-averaged PDFs $f_i = f_i(x,\mu_0^2)$ at the input scale are constructed according to
\begin{subequations}
\begin{align}
u &= u_v + 2 \bar{u}, \qquad
d  = d_v + 2 \bar{d},
\\         
\bar{u} &= S + \bar{u}_0, \hspace*{0.17cm} \qquad
\bar{d}  = S + \bar{d}_0,
\\
s       &= S + s_0 , \hspace*{0.20cm} \qquad \bar{s} = S + \bar{s}_0,
\end{align}
\end{subequations}
where each of the functions $u_v$, $d_v$, $S$, $\bar{u}_0$, $\bar{d}_0$, $s_0$ and $\bar{s}_0$, along with the gluon PDF $g(x,\mu_0^2)$, is modeled in terms of one template function in Eq.~(\ref{e.template}).
The valence PDFs $u_v = u - \bar{u}$ and $d_v = d - \bar{d}$ are isolated from the sea distributions in order to impose the quark number sum rules.
To ensure integrability of the first moment of each valence PDF, the corresponding $a_1$ parameters are restricted to be in the range $a_1 > -1$.
The parameters $a_3$ and $a_4$ are taken to be zero for $s_0$, $\bar{s}_0$ and $S$, but are free to vary for all other distributions.
The strangeness number sum rule 
    $\int_0^1  \dd{x} \pqty{s + \bar{s}} = 0$
imposes an additional constraint, which we use to fix the normalization of $s_0$.

The light sea quark distributions $\bar{u}$, $\bar{d}$, $s$ and $\bar{s}$ are modeled as combinations of a template function $S$ that dominates in the very small-$x$ region, with $a_1$ in the range $-2 < a_1 < -1$, and a template function for each of $\bar{u}_0$, $\bar{d}_0$, $s_0$ or $\bar{s}_0$ with $-1 < a_1 < 1$, controlling the shape at intermediate $x$ values.
The momentum sum rule is satisfied by adjusting the $a_0$ parameter of the gluon distribution.

For all template functions we restrict the $(1-x)$ exponent to be positive, $a_2 > 0$, to ensure that PDFs vanish in the limit $x \to 1$.
We choose the value $n = 2$ to normalize the $x$-dependent factor of the shape functions, since the momentum sum rule requires the existence of the second moment for each PDF flavor.
Other choices with $n > 2$ are also possible, and would be compensated by changes in the $a_0$ parameters.

For the spin-dependent PDFs $\Delta f_i = \Delta f_i(x,\mu_0^2)$ at the input scale we follow a similar strategy, but with a symmetric sea {\it ansatz}  due to the paucity of empirical constrains on the polarized sea quark distributions,
\begin{subequations}
\begin{align}
\Delta u &= \Delta u_v + 2 \Delta \bar{u},
\\
\Delta d &= \Delta d_v + 2 \Delta \bar{d},
\\         
\Delta \bar{u} &= \Delta \bar{d}
                = \Delta s
                = \Delta \bar{s}
                \equiv \Delta \bar{q},
\end{align}
\end{subequations}
where the functions $\Delta u_v$ and $\Delta d_v$ are modeled in terms of a template function in Eq.~(\ref{e.template}).
The sea quark helicity distribution $\Delta \bar{q}$ is described by a combination of a template function that dominates at very small $x$, with the $a_1$ parameter in the range $a_1 > -1$, and a second template function that controls the shape at intermediate $x$, with $a_1 > -0.5$, as in the unpolarized case.
For modelling the gluon helicity distribution we also adopt two template functions to allow sufficient flexibility for the fits.
In contrast to the spin-averaged case, we normalize the $x$-dependent factors for the spin-dependent template functions with $n = 1$, since all helicity PDFs are required to have finite contributions to the nucleon spin sum.
While the normalizations $a_0$ are in principle free parameters, as we will discuss in Sec.~\ref{ssec:addconstraints} below, we consider several different scenarios with either SU(2) or SU(3) flavor symmetry imposed, which provides additional constraints on the parameters.
For all the spin-dependent distributions, we fix the parameters $a_3$ and $a_4$ to zero.

In the next section, we present our data analysis framework for constraining the parameters of the spin-averaged and spin-dependent PDFs.
An important consequence of our simultaneous analysis will be the possibility to determine consistently the uncertainty quantification for the individual helicity-basis PDFs, defined in terms of the spin-averaged and spin-dependent distributions as
\begin{subequations}
\begin{align}
f^\uparrow   &= \frac12 \big(f + \Delta f\big),
\\
f^\downarrow &= \frac12 \big(f - \Delta f\big),
\end{align}
\label{e.helicitybasis}
\end{subequations}
for the helicity-aligned and antialigned distributions, respectively.

\section{Analysis framework}
\label{s.analysis}

The analysis framework adopted in this paper is based on the Bayesian Monte Carlo methodology developed in previous JAM analyses~\cite{Sato:2019yez, Moffat:2021dji, Cocuzza:2021rfn, Cocuzza:2021cbi}.
We also discuss additional constraints on the parameters arising from choices for the moments of the helicity PDFs respecting SU(2) or SU(3) flavor symmetry, as well as from positivity constraints on the PDFs (namely, both $f^\uparrow$ and $f^\downarrow$ being non-negative).
Before embarking on those discussion, however, we first review the experimental datasets that will be fitted in this analysis.

\subsection{Experimental datasets}
\label{s.data}

In the following we summarize the types and sources of experimental data for the observables listed in Table~\ref{tab:processes}, along with the kinematic cuts imposed in our analysis.

\subsubsection{Inclusive DIS}

The unpolarized fixed target inclusive DIS data included in our global fit are the reconstructed $F_2$ structure function data from BCDMS~\cite{Benvenuti:1989rh}, SLAC~\cite{Whitlow:1991uw}, and NMC~\cite{Arneodo:1996qe, Arneodo:1996kd}, with cuts
    $W^2 = M^2 + Q^2 (1-\xb)/\xb > 10$~GeV$^2$
and $Q^2 > m_c^2$, which are chosen to avoid power corrections and nuclear off-shell effects that are known to be more important at large values of $\xb$.
With the same cuts, we also include the reduced proton neutral current and charged current cross sections from the combined H1 and ZEUS analysis of HERA collider data~\cite{Abramowicz:2015mha}.

The spin-dependent DIS datasets include all fixed target experiments from the EMC~\cite{Ashman:1989ig}, SMC~\cite{Adeva:1998vv, Adeva:1999pa}, COMPASS~\cite{Alekseev:2010hc, Alexakhin:2006oza, Adolph:2015saz}, SLAC~\cite{Baum:1983ha, Abe:1998wq, Anthony:2000fn, Anthony:1999rm, Anthony:1996mw, Abe:1997cx} and HERMES~\cite{Ackerstaff:1997ws, Airapetian:2007mh} Collaborations, with identical cuts on $W^2$ and $Q^2$ as for the unpolarized DIS data~\cite{Sato:2016tuz, Ethier:2017zbq}.
Wherever possible, we use directly the experimental cross section asymmetries $A_{LL}=(\sigma^{++}-\sigma^{+-})/(\sigma^{++}+\sigma^{+-})$ rather than the reconstructed $g_1$ structure function.
This allows the consistent propagation of uncertainties on spin-dependent PDFs stemming from uncertainties in the unpolarized sector that enter through the denominators of the polarization asymmetries.
A total of 2680 unpolarized and 365 polarized DIS data points are used in the analysis.

\subsubsection{Drell-Yan lepton-pair production}

In addition to inclusive DIS data, we also fit Drell-Yan inclusive lepton-pair production data from $pp$ and $pD$ collisions, available from the Fermilab E866 experiment~\cite{Hawker:1998ty}.
The data are selected to exclude regions of invariant mass $M_{\ell^+\ell^-}$
of the lepton-pair in the vicinity of the $q \bar{q}$ resonances, such as the $J/\psi$, $\psi'$ and $\Upsilon$ states.
Furthermore, following Alekhin {\it et~al.} \cite{Alekhin:2006zm}, we also apply a cut of $M_{\ell^+\ell^-} > 6$~GeV to avoid tension with DIS data~\cite{Alekhin:2006zm}, which leaves 250 data points available to be fitted.

\subsubsection{Inclusive jet production}

Inclusive jet production data are known for providing unique sensitivity to the gluonic content of hadrons.
For unpolarized beams, we use existing $p \overline{p}$ data from the D0~\cite{Abazov:2008ae} and CDF \cite{Abulencia:2007ez} Collaborations at the Tevatron.
In addition, we include for the first time in a global fit the jet production data in $pp$ collisions from the STAR Collaboration at RHIC~\cite{Abelev:2006uq}.
For the latter, we include data from the 2003 and 2004 runs, and find that they can be described well for jet transverse momenta $p_T > 8$~GeV.

With polarized proton beams, we use double spin asymmetries $A_{LL}$ from the STAR \cite{Abelev:2006uq, Adamczyk:2012qj, Adamczyk:2014ozi, Adam:2019aml, STAR:2021mfd, STAR:2021mqa} and PHENIX \cite{Adare:2010cc} Collaborations at RHIC.
We restrict the data to the same $p_T$ range as for the unpolarized jet cross sections in order to guarantee a faithful description of the denominator in the asymmetries.
For the renormalization and factorization scales $\mu_R$ and $\mu_F$, respectively, we generally find best agreement for both unpolarized and polarized collisions with a scale $p_T/2$.

The treatment of systematic uncertainties of the jet data used in our analysis requires particular attention.
For the CDF data, uncertainties are provided that are only correlated within each rapidity bin and have to be treated separately.
Furthermore, both D0~\cite{Abazov:2008ae} and CDF~\cite{Abulencia:2007ez} provide parton to hadron correction factors obtained from Monte Carlo simulations~\cite{Abulencia:2007ez}, which translate the parton level calculation to the hadron level, and these are incorporated in our analysis.

For the STAR jet data, an uncertainty in the relative luminosity measurement usually results in a shift of the $A_{LL}$ data by an additive constant (fully correlated systemic uncertainty)~\cite{Adamczyk:2012qj}, while an uncertainty in the measurement of the polarization magnitude scales the $A_{LL}$ data (normalization uncertainty).
We implement these uncertainties in our analysis as described in Sec.~\ref{s.bayes} below.
To avoid underestimating uncertainties, we treat weakly correlated uncertainties as uncorrelated, in particular, for the STAR jet $A_{LL}$ data from the 2005 \cite{Adamczyk:2012qj}, 2009 \cite{Adamczyk:2014ozi} and 2013 \cite{STAR:2021mqa} and 2015 \cite{STAR:2021mfd} runs.

\begin{figure}[t]
\includegraphics[width = 0.6\textwidth]{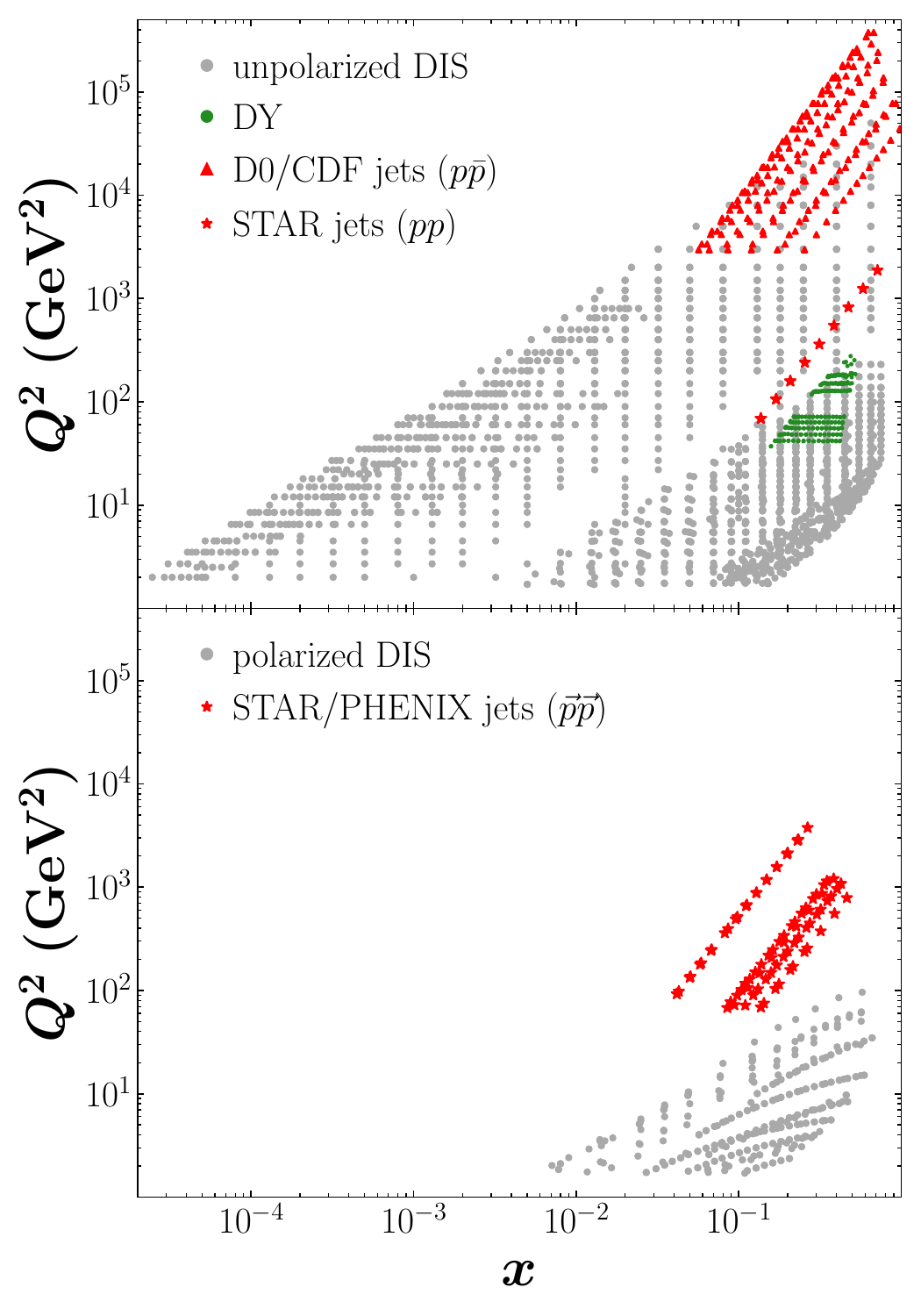}
\caption
{
Kinematic coverage of the unpolarized (upper panel) and polarized (lower panel) datasets used in this analysis.
The unpolarized datasets include fixed-target and HERA collider DIS data (gray solid circles), Drell-Yan from $pp$ and $pD$ collisions at Fermilab (green solid circles), and jet production from unpolarized $p \overline{p}$ scattering at Tevatron (red upward triangles) and $pp$ scattering at STAR (red stars).
The polarized datasets include spin-dependent fixed-target DIS data (gray solid circles), and jet production in polarized $pp$ scattering at STAR and PHENIX from RHIC (red stars).
The variable $x$ denotes the Bjorken scaling variable $\xb$ for DIS and the Feynman variable $\xf$ for Drell-Yan and jet production, while the scale $Q^2$ represents the four-momentum transfer squared for DIS and DY, and transverse momentum squared for jets.
}
\label{f.kin}
\end{figure}

The global kinematic coverage of the unpolarized and polarized datasets is shown in \cref{f.kin}.
Clearly, the range of kinematics spanned by the unpolarized data is significantly greater than that of the polarized (by some 2 orders of magnitude at small $x$ and 2 orders of magnitude at high $Q^2$).
Importantly, however, the range covered by the polarized data overlaps with the unpolarized, so that in this range a consistent analysis of spin-dependent and spin-averaged PDFs can be achieved.

\subsection{Bayesian inference}
\label{s.bayes}

Our analysis framework is based on sampling the Bayesian posterior distribution using the data resampling approach.
The posterior distribution is given by a product of the likelihood function ${\cal L}$ and a prior distribution $\pi$,
\begin{equation}
\label{e.bayes}
\rho(\boldsymbol{a},\boldsymbol{\nu}\,\big|\,\mathrm{data}) 
    \propto 
    \mathcal{L}(\boldsymbol{a},\boldsymbol{\nu}\,|\,\mathrm{data})\, \pi(\boldsymbol{a},\boldsymbol{\nu}), 
\end{equation}
where we distinguish the PDF parameters $\boldsymbol{a}$ from the additional nuisance parameters $\boldsymbol{\nu}$ (see below).
For the likelihood function we use a Gaussian of the form 
\begin{equation}
\mathcal{L} (\boldsymbol{a},\boldsymbol{\nu}\,|\,\mathrm{data})
= \exp \bigg[
-\frac12 
\sum_{e,i} 
\bigg(
\frac{d_{e,i} - t_{e,i}(\boldsymbol{a},\boldsymbol{\nu})}{\alpha_{e,i}}
\bigg)^{\!2}\,
\bigg],
\end{equation}
where $d_{e,i}$ is the $i$-th data point from experiment $e$ with uncorrelated uncertainty $\alpha_{e,i}$, and $t_{e,i}$ is the corresponding theoretical value.
Since the experimental data are distorted by detector effects, we model such distortion with a multiplicative normalization parameter and additive shifts within the quoted systematic uncertainties.
Specifically, we compute each theoretical value $t_{e,i}$ according to
\begin{align}
    t_{e,i}(\boldsymbol{a},\boldsymbol{\nu})\,
    =\, \sum_k r_{e,k}\, \beta_{e, k, i}\,
    +\, \frac{1}{N_e}\, t^0_{e,i}(\boldsymbol{a}),
\end{align}
with nuisance parameters $\boldsymbol{\nu} = \{ r_{e,k}; N_e \}$, where $k$ labels different types of systematic uncertainties for the $i$-th data point of the experiment $e$. 
The theoretical calculation of the observable $t^0_{e,i}$ is obtained by convoluting the PDFs with the short-distance cross sections, which is then multiplied by a nuisance parameter $N_e$.
The additive shifts are controlled by the nuisance parameters $r_{e,k}$, and weighted by a set of quoted point-by-point correlated systematic uncertainties $\beta_{e,k,i}$.

The prior distribution $\pi(\boldsymbol{a},\boldsymbol{\nu})$ includes flat priors for the PDF parameters $a_i \in [a_i^{\rm min}, a_i^{\rm max}]$ and $\delta$ functions to impose the valence quark number and momentum sum rules in the case of unpolarized PDFs.
The priors for the nuisance parameters are included as Gaussian penalties.
For the normalization parameters, the corresponding priors are modeled in a Gaussian form using the quoted experimental normalization uncertainty, $\delta N_e$.
The full prior distribution can then be explicitly written as
\begin{align}
    \pi(\boldsymbol{a},\boldsymbol{\nu})
   = & ~\delta\big(a^{u_v}_0 - \mathrm{SR} (\boldsymbol{a}^{u_v} , u_v)\big)
       ~\delta\big(a^{d_v}_0 - \mathrm{SR} (\boldsymbol{a}^{d_v} , d_v)\big)
       ~\delta\big(a^{s}_0   - \mathrm{SR} (\boldsymbol{a}^{s} , s)\big)
       ~\delta\big(a^g_0 - \mathrm{SR} (\boldsymbol{a}^g , g)\big)
\notag\\
    & \times
    ~\prod_i \theta\big(a^{\min}_i<a_i<a^{\max}_i\big)
    ~\prod_e\prod_k \exp\Big(\!\!-\!\frac12 r_{e,k}^2 \Big)\,
             \exp\Big(\!\!-\!\frac{1-N_e}{2\, \delta N_e} \Big),
\label{e.priors}
\end{align}
where ``SR'' represents instructions to adjust the $a_0$ parameters in order to satisfy the appropriate sum rules~\cite{zhou:thesis}, and the $\delta$ functions are implemented analytically.

In adopting the data resampling approach to the construction of Monte Carlo samples for the posterior distribution, we add Gaussian noise to each experimental data point, $d_{e,i} \to d_{e,i} + R_{e,i}\, \alpha_{e,i}$, with statistically-independent normally distributed random numbers $R_{e,i}$.
With a given set of distorted data, an optimization is performed in parameter space that maximizes the posterior distribution $\rho(\boldsymbol{a},\boldsymbol{\nu}\,|\,\mathrm{data}\!+\!{\rm noise})$.
The resulting parameters are added to the list of Monte Carlo samples and the process is repeated $O(1,000)$ times in order to accumulate sufficient samples to compute statistical estimators for a given observable ${\cal O}$
\begin{subequations}
\begin{align}
\mathrm{E} [\mathcal{O}]
& = \frac{1}{N} \sum_k \mathcal{O} (\boldsymbol{a}_k),
\label{e.Edef}
\\
\mathrm{V} [\mathcal{O}]
& = \frac{1}{N} \sum_k \Big[ \mathcal{O} (\boldsymbol{a}_k) - \mathrm{E}[\mathcal{O}] \Big]^2,
\label{e.Vdef}
\end{align}
\label{e.EandVdef}
\end{subequations}
where $N$ is the number of parameter samples drawn from the posterior distribution.

While in principle each optimization in the sampling procedure can be carried out with all parameters open, in practice exploring a highly dimensional parameter space (which in our case is $\approx 80$, including 30 parameters for spin-averaged PDFs, 18 for spin-dependent PDFs, and 33 normalizations) becomes inefficient due to local minima or vanishing gradients, and only a handful of samples can be collected.
This can be overcome by using the multi-step methodology developed in Ref.~\cite{Sato:2019yez}, where the initial parameters for the optimization are pre-tuned by using a restricted set of data.
Fewer datasets decreases the numerical expense of drawing parameter samples from the posterior via the data resampling, and the procedure can be repeated in several steps with different datasets added sequentially.
The resulting samples from a given step become the input samples for the next step, where additional datasets are added.

In addition, since the spin-dependent PDFs do not enter in unpolarized physical observables, our sequence of pre-tuning the PDFs parameters can be partitioned into three stages.
In the first stage only the unpolarized PDF parameters are tuned.
In the second stage the Monte Carlo parameters samples of the unpolarized PDFs are frozen and used to tune the spin-dependent PDF parameters.
The final stage uses the collection of pre-optmized input parameters for all PDFs, and performs a final posterior sampling with all parameters free and all datasets included in the posterior distribution.
A sketch of the multi-step strategy for the parameter sampling is illustrated in \fref{multistep}.
Finally, for our diagnostic metric for the global agreement between data and theory, we use the reduced $\chi^2$, defined as
\begin{align}
\chi^2_{\rm red} 
    \equiv \frac{\chi^2}{N_{\rm dat}} 
    = \frac{1}{N_{\rm dat}} 
        \sum_{e,i} \left(
            \frac{d_{e,i} - {\rm E}[t_{e,i}]}
                {\alpha_{e,i}}
            \right)^2,
\label{e.chi2dof}
\end{align}
where the expectation value ${\rm E}[t_{e,i}]$ for the theoretical quantity $t_{e,i}$ is estimated via \eref{Edef}.

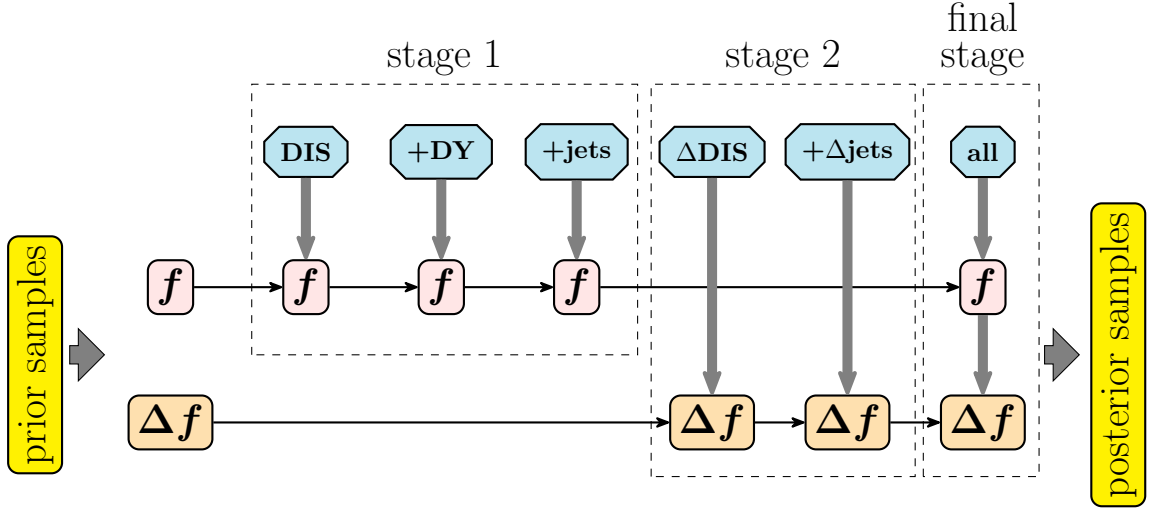
\begin{figure}[t]
\begin{center}
\begin{tikzpicture}[transform shape, scale = 0.9]
\draw (0.0, 0.0) node[rectangle, thick, rounded corners, draw = black, fill = yellow, rotate = 90] (prior) {\Large{prior samples}};

\draw (2.0, 1.0) node[rectangle, thick, rounded corners, draw = black, fill = pink!40.0] (pdf_0) {\Large $\boldsymbol{f}$};
\draw (2.0, -1.0) node[rectangle, thick, rounded corners, draw = black, fill = Dandelion!40.0] (ppdf_0) {\Large $\boldsymbol{\Delta f}$};

\draw (4.0, 1.0) node[rectangle, thick, rounded corners, draw = black, fill = pink!40.0] (pdf_1) {\Large $\boldsymbol{f}$};
\draw (4.0, 3.0) node[chamfered rectangle, thick, draw = black, fill = SkyBlue!40.0] (dis) {\textbf{DIS}};

\draw (6.0, 1.0) node[rectangle, thick, rounded corners, draw = black, fill = pink!40.0] (pdf_2) {\Large $\boldsymbol{f}$};
\draw (6.0, 3.0) node[chamfered rectangle, thick, draw = black, fill = SkyBlue!40.0] (dy) {\textbf{+DY}};

\draw (8.0, 1.0) node[rectangle, thick, rounded corners, draw = black, fill = pink!40.0] (pdf_3) {\Large $\boldsymbol{f}$};
\draw (8.0, 3.0) node[chamfered rectangle, thick, draw = black, fill = SkyBlue!40.0] (jet) {\textbf{+jets}};

\draw (10.0, -1.0) node[rectangle, thick, rounded corners, draw = black, fill = Dandelion!40.0] (ppdf_1) {\Large $\boldsymbol{\Delta f}$};
\draw (10.0, 3.0) node[chamfered rectangle, thick, draw = black, fill = SkyBlue!40.0] (pdis) {\textbf{$\Delta$DIS}};

\draw (12.0, -1.0) node[rectangle, thick, rounded corners, draw = black, fill = Dandelion!40.0] (ppdf_2) {\Large $\boldsymbol{\Delta f}$};
\draw (12.0, 3.0) node[chamfered rectangle, thick, draw = black, fill = SkyBlue!40.0] (pjet) {\textbf{+$\Delta$jets}};

\draw (14.0, 1.0) node[rectangle, thick, rounded corners, draw = black, fill = pink!40.0] (pdf_last) {\Large $\boldsymbol{f}$};
\draw (14.0, -1.0) node[rectangle, thick, rounded corners, draw = black, fill = Dandelion!40.0] (ppdf_last) {\Large $\boldsymbol{\Delta f}$};
\draw (14.0, 3.0) node[chamfered rectangle, thick, draw = black, fill = SkyBlue!40.0] (all) {\textbf{all}};

\draw (16.0, 0.0) node[rectangle, thick, rounded corners, draw = black, fill = yellow, rotate = 90] (posterior) {\Large{posterior samples}};

\draw [-{Stealth[round]}, thick] (pdf_0.east) -- (pdf_1.west);
\draw [-{Stealth[round]}, thick] (pdf_1.east) -- (pdf_2.west);
\draw [-{Stealth[round]}, thick] (pdf_2.east) -- (pdf_3.west);
\draw [-{Stealth[round]}, thick] (pdf_3.east) -- (pdf_last.west);

\draw [-{Stealth[round]}, thick] (ppdf_0.east) -- (ppdf_1.west);
\draw [-{Stealth[round]}, thick] (ppdf_1.east) -- (ppdf_2.west);
\draw [-{Stealth[round]}, thick] (ppdf_2.east) -- (ppdf_last.west);

\draw [-{Stealth[round]}, double = gray, ultra thick, draw = gray] (dis.south) -- (pdf_1.north);
\draw [-{Stealth[round]}, double = gray, ultra thick, draw = gray] (dy.south) -- (pdf_2.north);
\draw [-{Stealth[round]}, double = gray, ultra thick, draw = gray] (jet.south) -- (pdf_3.north);

\draw [-{Stealth[round]}, double = gray, ultra thick, draw = gray] (pdis.south) -- (ppdf_1.north);
\draw [-{Stealth[round]}, double = gray, ultra thick, draw = gray] (pjet.south) -- (ppdf_2.north);
\draw [-{Stealth[round]}, double = gray, ultra thick, draw = gray] (all.south) -- (pdf_last.north);
\draw [-{Stealth[round]}, double = gray, ultra thick, draw = gray] (pdf_last.south) -- (ppdf_last.north);

\node[single arrow, draw, fill = gray] at (0.7, 0.0) {\thickspace};
\node[single arrow, draw, fill = gray] at (15.1, 0.0) {\thickspace};

\draw[dashed] (3.2, 0.0) rectangle (8.9, 4.0);
\draw (6.05, 4.0) node[above] {\Large{stage 1}};
\draw[dashed] (9.1, -1.8) rectangle (13.0, 4.0);
\draw (11.05, 4.0) node[above] {\Large{stage 2}};
\draw[dashed] (13.12, -1.8) rectangle (14.83, 4.0);
\draw (14.0, 4.65) node[above] {\Large{final}};
\draw (14.0, 4.0) node[above] {\Large{stage}};
\end{tikzpicture}
\end{center}
\caption
[Multi-step strategy]
{Schematic overview of the multi-step strategy for parameter sampling employed in the current JAM analysis of spin-averaged ($f$) and spin-dependent ($\Delta f$) PDFs.  Unpolarized data include DIS, Drell-Yan (DY), and single jet production in $pp$ and $p\bar p$ collisions, while polarized data include polarized DIS ($\Delta$DIS) and jet production in polarized $pp$ collisions ($\Delta$jets).}
\label{f.multistep}
\end{figure}

\subsection{Additional constraints}
\label{ssec:addconstraints}

In contrast to the spin-averaged PDFs, where the valence number and momentum sum rules can be imposed based on fundamental physical properties such as baryon number and momentum conservation, no correspondings constraints are available for spin-dependent PDFs.
On the other hand, approximate sum rules for spin-dependent PDFs exist involving the triplet and octet axial-vector charges, $g_A$ and $a_8$, respectively,
\begin{subequations}
\begin{align}
& \int_0^1 \dd{x}
\big[ \Delta u^+ - \Delta d^+ \big](x,Q^2)
= g_A,
\label{e.su_2} 
\\
& \int_0^1 \dd{x}
\big[ \Delta u^+ + \Delta d^+ - 2 \Delta s^+ \big](x,Q^2)
= a_8, 
\label{e.su_3}
\end{align}
\label{e.ac}%
\end{subequations}%
where $\Delta q^+ = \Delta q +\Delta \bar{q}$.
These relations can be imposed as constraints if the values of $g_A$ and $a_8$ are known sufficiently accurately.
In practice, however, the charges are inferred from neutron beta-decays using SU(2) isospin symmetry, which gives
    $g_A = 1.269(3)$,
and from hyperon beta-decays, which under the assumption of SU(3) flavor symmetry gives
    $a_8 = 0.586(31)$~\cite{Ethier:2017zbq}.
%
While these constraints are typically imposed in global analyses of spin-dependent PDFs~\cite{deFlorian:2014yva, Nocera:2014gqa, Sato:2016tuz}, in our analysis we consider several scenarios where one or both of these are imposed or relaxed.
Imposing Eqs.~(\ref{e.su_2}) or (\ref{e.su_3}) amounts to extending the priors in \eref{priors} according to
\begin{align}
    \pi(\boldsymbol{a},\boldsymbol{\nu})\
    \to\ \pi(\boldsymbol{a},\boldsymbol{\nu})
    &\times 
    \exp\bigg(\!-\!\frac{a^{\rm exp}_3-a^{\rm thy}_3}{2\, \delta a^{\rm exp}_3}\bigg)
\hspace*{2.18cm} {\rm [SU(2)]}
\notag\\
    &\times
    \exp\bigg(\!-\!\frac{a^{\rm exp}_8-a^{\rm thy}_8}{2\, \delta a^{\rm exp}_8}\bigg) ,
\hspace*{2cm} {\rm [SU(3)]}
\end{align}
where the theoretical values $a_{3,8}^{\rm thy}$ are computed using Eqs.~(\ref{e.ac}) and the experimental values $a_{3,8}^{\rm exp}$ and $\delta a_{3,8}^{\rm exp}$ are taken from Ref.~\cite{Ethier:2017zbq}.

In addition to the SU(2) and SU(3) flavor symmetry constraints on the axial-vector charges, many global analyses also impose phenomenological positivity constraints on the PDFs \cite{deFlorian:2014yva, Nocera:2014gqa}, which requires that 
\begin{equation}
\hspace*{2cm} 
\big|\Delta q(x,Q^2)\big| 
\leqslant q(x,Q^2),
\hspace*{3cm} {\rm [positivity]}
\label{e.positivity}
\end{equation}
for each PDF flavor $q$, or equivalently that all the helicity-basis PDFs remain positive, $f^{\uparrow/\downarrow}(x,Q^2) \geqslant 0$.
The positivity constraints typically affect PDFs in the large-$x$ region, where the absolute magnitudes of the spin-dependent and spin-averaged PDFs become comparable, and can be implemented by selecting the posterior samples that satisfy such criteria.
On the other hand, it has been recently argued~\cite{Collins:2021vke}, in contrast to earlier claims in the literature~\cite{Candido:2020yat}, that there is no fundamental requirement for PDFs in the $\overline{\rm MS}$ scheme to remain positive definite.

Whether formally justified or not, it is important to assess the degree to which the constraints (\ref{e.ac}) and (\ref{e.positivity}) may bias the inference on the PDFs.
To this effect, we consider three scenarios, in which we impose either
\begin{enumerate}
\item[(i)]  the {\bf SU(2)} constraint in Eq.~(\ref{e.su_2});
\item[(ii)] the {\bf SU(3)} constraint in Eq.~(\ref{e.su_3}) in addition to (\ref{e.su_2}); or
\item[(iii)] the {\bf SU(3) + positivity} constraints in Eq.~(\ref{e.positivity}) in addition to Eqs.~(\ref{e.su_2}) and (\ref{e.su_3}).
\end{enumerate}
In the next section we analyze in detail the effect on the global QCD analysis of each of these assumptions.

\section{Simultaneous analysis results}
\label{s.results}

Using the theoretical framework and methodology outlined in Secs.~\ref{s.theory} and \ref{s.analysis}, in this section we present the results of our global Monte Carlo analysis, simultaneously extracting both the spin-averaged and spin-dependent PDFs.
We begin with the discussion of the spin-averaged PDFs, and in particular the impact of jet data from unpolarized $pp$ collisions at RHIC, obtained at similar kinematics as the polarized $pp$ jet data that are expected to constrain the spin-dependent gluon PDF, $\Delta g$.
We focus in particular on the determination of $\Delta g$ in the context of the three scenarios discussed in the previous section, and critically examine the discrimination between the two helicity-basis PDFs under these scenarios.
A summary of the fit results, including the $\chi_{\rm red}^2$ values for each type of dataset and for each of the scenarios, is given in \cref{t.chi_2}. 
The $\chi^2_{\rm red}$ are computed from Eq.~(\ref{e.chi2dof}) using the average of theory predictions from all replicas in the simultaneous fit.
%

\begin{table}[b]
\caption
[$\chi ^2$ for JAM fits]
{ 
Results for the simultaneous fits to unpolarized and polarized scattering data, with the reduced $\chi^2_{\rm red} = \chi^2/N_{\rm dat}$ for $N_{\rm dat}$ points for the SU(2), SU(3), and SU(3)+positivity scenarios.
The $\chi_{\mathrm{red}}^2$ values computed from all the samples are shown, along with corresponding values from the positive (superscript) and negative (subscript) $\Delta g$ solutions.
}
\begin{tabular}{l|r|ccc}
\hline
\multirow{2}{*}{~Data} & \multirow{2}{*}{$N_\mathrm{dat}$~~} & \multicolumn{3}{c}{$\chi^2_{\rm red}$} \\
\cline{3-5}
& & ~~~\footnotesize{\bf SU(2)}~~ 
    & ~\footnotesize{\bf SU(3)}~ 
    & \footnotesize{\bf SU(3)+pos} \\
\hline
~Unpolarized DIS~\cite{Benvenuti:1989rh, Whitlow:1991uw, Arneodo:1996qe, Arneodo:1996kd, Abramowicz:2015mha}                                     &~2680~ 
& ~~$1.20_{\,(1.20)}^{\,(1.20)}$~~  & ~~$1.20_{\,(1.21)}^{\,(1.20)}$~~  & 1.20      \\
~Drell-Yan ($pp$, $pD$) \cite{Hawker:1998ty}             &  250~
& $1.06_{\,(1.06)}^{\,(1.05)}$  & $1.06_{\,(1.06)}^{\,(1.06)}$  & 1.10 \\ 
~Jets
&      &      &      &      \\
\qquad D0 ($p \overline{p}$) \cite{Abazov:2008ae}~       &  110~
& $0.89_{\,(0.89)}^{\,(0.89)}$  & $0.89_{\,(0.89)}^{\,(0.89)}$  & 0.88      \\
\qquad CDF ($p \overline{p}$) \cite{Abulencia:2007ez}~   &   76~
& $1.11_{\,(1.11)}^{\,(1.11)}$  & $1.11_{\,(1.11)}^{\,(1.11)}$  & 1.11      \\
\qquad STAR 2003 ($p p$) \cite{Abelev:2006uq}~           &    3~
& $0.04_{\,(0.04)}^{\,(0.04)}$  & $0.04_{\,(0.04)}^{\,(0.04)}$  & 0.04      \\
\qquad STAR 2004 ($p p$) \cite{Abelev:2006uq}~           &    9~
& $1.06_{\,(1.05)}^{\,(1.06)}$  & $1.06_{\,(1.05)}^{\,(1.06)}$  & 1.06      \\
\hline
~Polarized DIS~\cite{Ashman:1989ig, Adeva:1998vv, Adeva:1999pa, Alekseev:2010hc, Alexakhin:2006oza, Adolph:2015saz, Baum:1983ha, Abe:1998wq, Anthony:2000fn, Anthony:1999rm, Anthony:1996mw, Abe:1997cx, Ackerstaff:1997ws, Airapetian:2007mh}~
                                                         & ~365~
& $0.92_{\,(0.94)}^{\,(0.92)}$  & $0.92_{\,(0.95)}^{\,(0.92)}$  & 0.96      \\
~Jets in polarized $\vec{\,p} \vec{\,p}$  &
&       &       &           \\
\qquad STAR \cite{Abelev:2006uq, Adamczyk:2012qj, Adamczyk:2014ozi, Adam:2019aml, STAR:2021mfd, STAR:2021mqa}~                            &  81~
& $0.82_{\,(0.86)}^{\,(0.83)}$  & $0.81_{\,(0.85)}^{\,(0.82)}$  & 0.84      \\
\qquad PHENIX \cite{Adare:2010cc}                        &   2~
& $0.38_{\,(0.39)}^{\,(0.38)}$  & $0.38_{\,(0.39)}^{\,(0.38)}$  & 0.38      \\
\hline
{\bf ~Total}                                             & ~{\bf 3576}~
& {\bf 1.14}    & {\bf 1.14}    & {\bf 1.15}    \\
\hline
\end{tabular}
\label{t.chi_2}
\end{table}

\subsection{Unpolarized PDFs}

\begin{figure}[t]
\begin{center}
\includegraphics[width = 0.99 \textwidth]{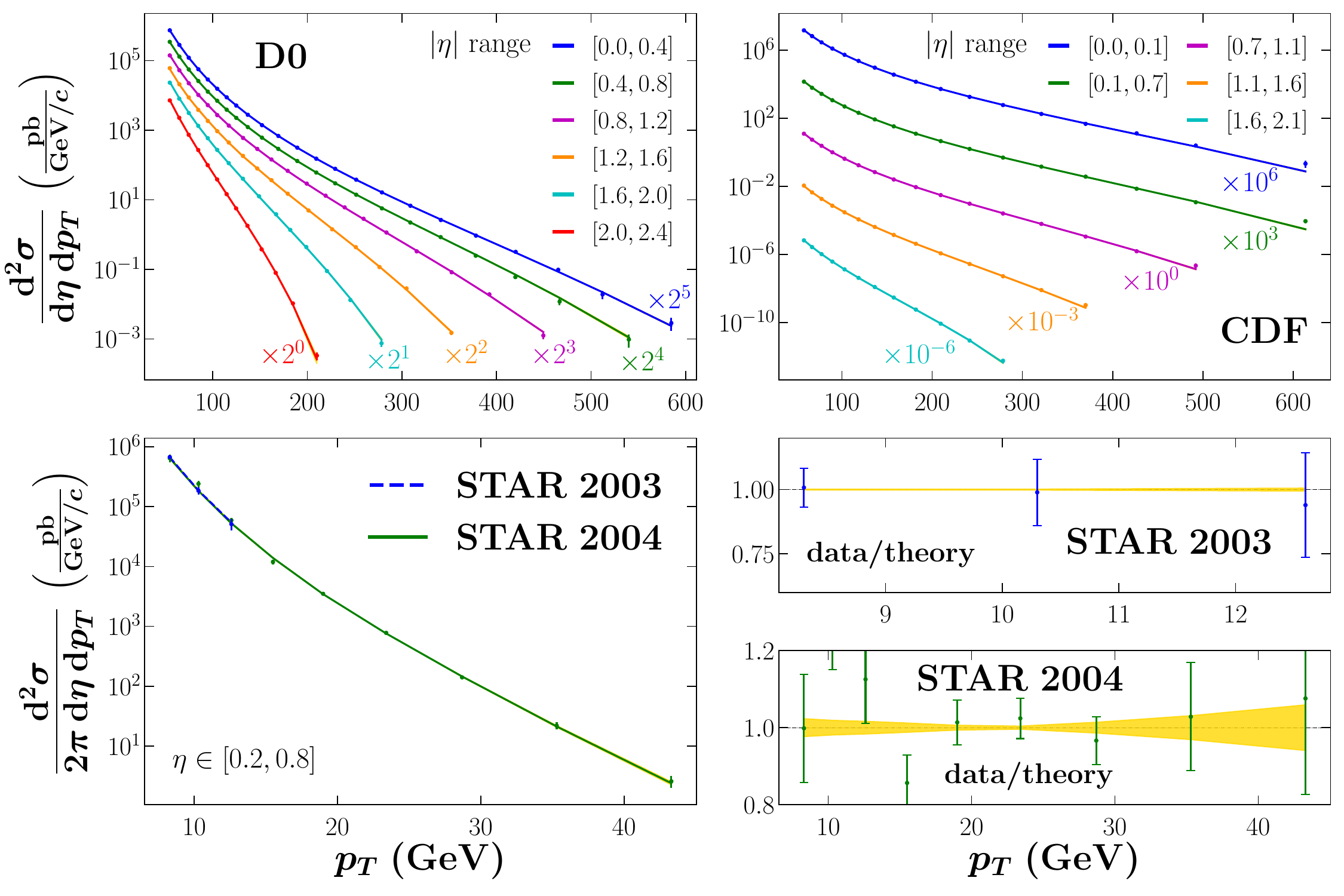}
\caption
[Fit to unpolarized jet observables from Tevatron and RHIC]
{Comparison with single jet production cross sections in $p\bar p$ collisions from D0 \cite{Abazov:2008ae} and CDF \cite{Abulencia:2007ez}, and in $pp$ scattering from STAR \cite{Abelev:2006uq}. Different pseudorapidity $\eta$ bins are marked by colors and scaled by factors for D0 and CDF for clarity. Note the extra factor $2 \pi$ in the STAR cross section data. The data (filled circles) are compared with fits (solid lines) obtained from the average of all Monte Carlo replicas, with $1\sigma$ uncertainties (yellow bands). For STAR 2003 and 2004 data, ratios of data to average theory (filled circles) are shown in the bottom right panel.}
\label{f.jet_fit}
\end{center}
\end{figure}

As indicated in \cref{t.chi_2}, good overall agreement is found between our fits and the unpolarized DIS (fixed target and HERA collider), Drell-Yan, and inclusive jet production data.
Since the focus of this work is primarily on jet observables (the full fit results for the inclusive DIS and Drell-Yan data comparisons can be found in Ref.~\cite{zhou:thesis}), in \cref{f.jet_fit} we show the differential jet production cross sections,
    $\dd^2{\sigma}/\dd{\eta}\dd{p_T}$,
for $p\bar p$ scattering from D0 and CDF at the Tevatron and $pp$ scattering from STAR at RHIC versus the jet transverse momentum $p_T$, in specific bins of the pseudorapidity $\eta$.
The $\eta$ bins are obtained using their absolute values for CDF and D0 data, and actual values for STAR.
An excellent description of all the jet data is obtained, with $\chi^2_{\rm red} \approx 0.9 - 1.1$ for the D0 and CDF $p\bar p$ data, and the STAR 2004 $pp$ data.
Since only 3 data points are available for the STAR 2003 run, a $\chi^2_{\rm red} \approx 0$ was obtained for these points.
The $\chi^2_{\rm red}$ values for the unpolarized data are, as may be expected, almost independent of the scenario chosen for the spin PDF constraints, Eqs.~(\ref{e.ac}) and (\ref{e.positivity}).

To more graphically illustrate the comparison between data and theory for the STAR $pp$ data, which have not been used in any previous global QCD analysis, we also show in \cref{f.jet_fit} the data/theory ratios, which our fit describes well within the $\sim 10\%-20\%$ experimental uncertainties.
Note that the cross sections vary by some 7 orders of magnitude for the D0 and CDF data over the range of $p_T$ covered ($p_T \lesssim 600$~GeV), and over 5 orders of magnitude for the STAR data, which span a smaller range of $p_T$ values ($p_T \lesssim 40$~GeV).
In principle, the data are available down to rather low $p_T$ values, $p_T \sim$ few GeV. 
However, some tensions were found when attempting to fit the STAR data with $p_T < 8$~GeV and those with $p_T > 8$~GeV, so that a cut of $p_T > 8$~GeV is made for this analysis.
This will have some consequence for the corresponding cut chosen for the polarized scattering data, as we discuss below.

\begin{figure}[t]
\begin{center}
\includegraphics[width = 0.9 \textwidth]{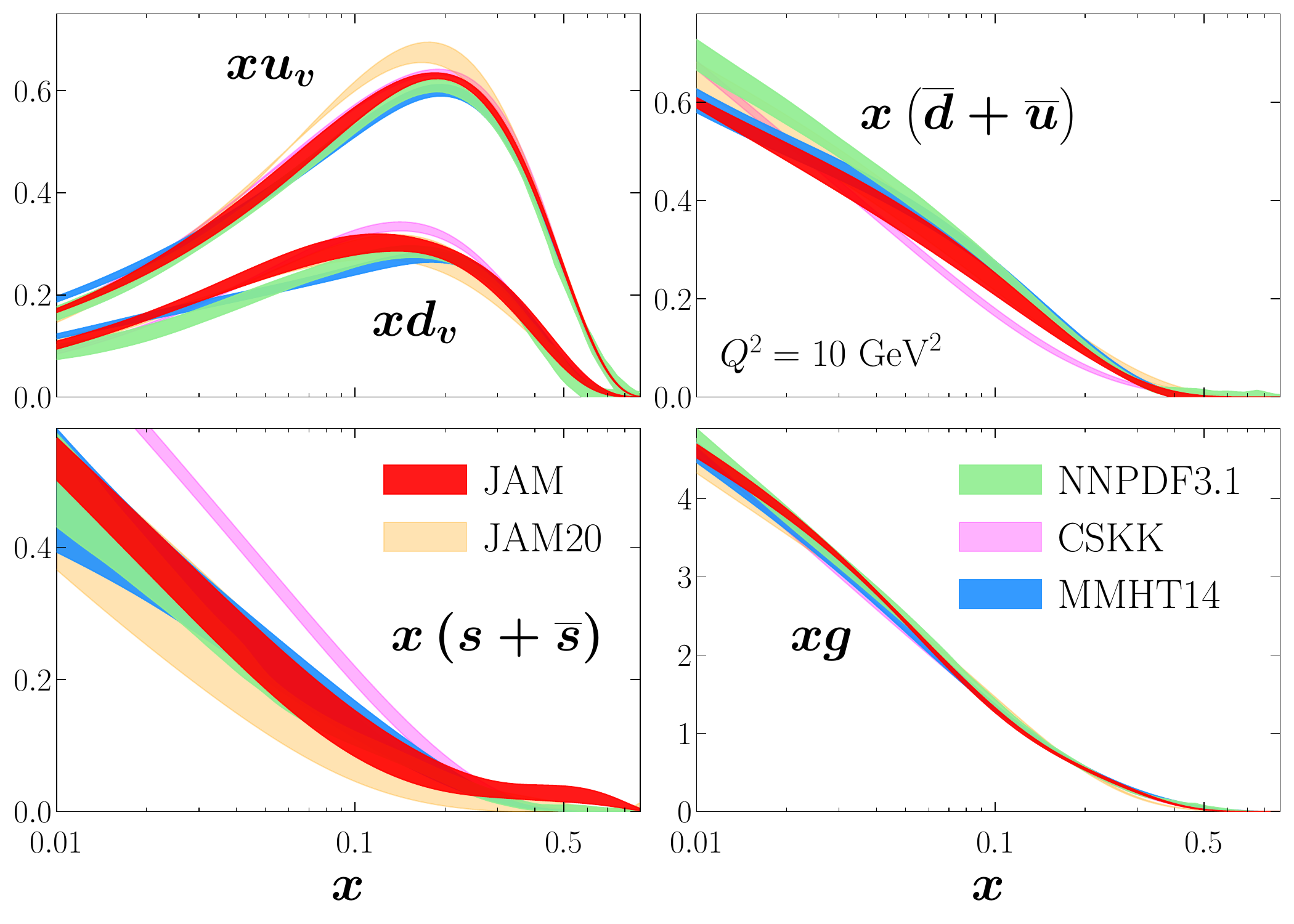}
\caption
[Comparison of spin-averaged PDFs with other groups]
{
Comparison of spin-averaged PDFs from the present JAM analysis with other PDF sets from the previous JAM20 analysis~\cite{Sato:2019yez}, and from the NNPDF3.1 \cite{NNPDF:2017mvq}, CSKK \cite{Cooper-Sarkar:2018ufj}, and MMHT14 \cite{Harland-Lang:2014zoa} parametrizations, for the $u_v$, $d_v$, $\bar d+\bar u$, and $s+\bar s$ quark and gluon $g$ flavors at $Q^2 = 10$~GeV$^2$. Note that $x$ times the PDF is shown.
}
\label{f.pdf_groups}
\end{center}
\end{figure}

The unpolarized PDFs extracted from the present analysis are shown in \cref{f.pdf_groups} for the valence quark $u_v$ and $d_v$, light antiquark $\bar d+\bar u$, and strange $s+\bar s$ distributions, as well as the gluon PDF, $g$, at a scale $Q^2=10$~GeV$^2$.
Compared with PDFs from several other global analyses~\cite{Sato:2019yez, NNPDF:2017mvq, Cooper-Sarkar:2018ufj, Harland-Lang:2014zoa}, the variation between the different PDF sets is relatively small for the valence, light antiquark, and gluon PDFs, while a larger spread is observed for the strange quark distributions.
In particular, the magnitude of the strange $s+\bar s$ PDFs is slightly larger than the earlier JAM20~\cite{Moffat:2021dji} (and also JAM19~\cite{Sato:2019yez}, not shown in \cref{f.pdf_groups}) analysis, which found a stronger strange quark suppression due to the inclusion of semi-inclusive DIS and single-inclusive $e^+ e^-$ annihilation data, especially for kaon production.
We expect that inclusion of the semi-inclusive DIS and $e^+ e^-$ annihilation data into the present analysis will produce additional suppression of our strange quark PDF.
Also, at higher $x$ values ($x \gtrsim 0.5$) the upward shift in the strangeness PDF is an indirect effect associated with the sensitivity of the jet data to the gluon PDF via the momentum sum rule.
However, the detailed structure of the strange quark PDFs does not affect the main goal of our analysis, which is the determination of the spin-dependent gluon distribution.

%

\subsection{Spin-dependent PDFs}
\label{ss.spin_pdfs}

\begin{figure}[t]
\begin{center}
\includegraphics[width = 0.95 \textwidth]{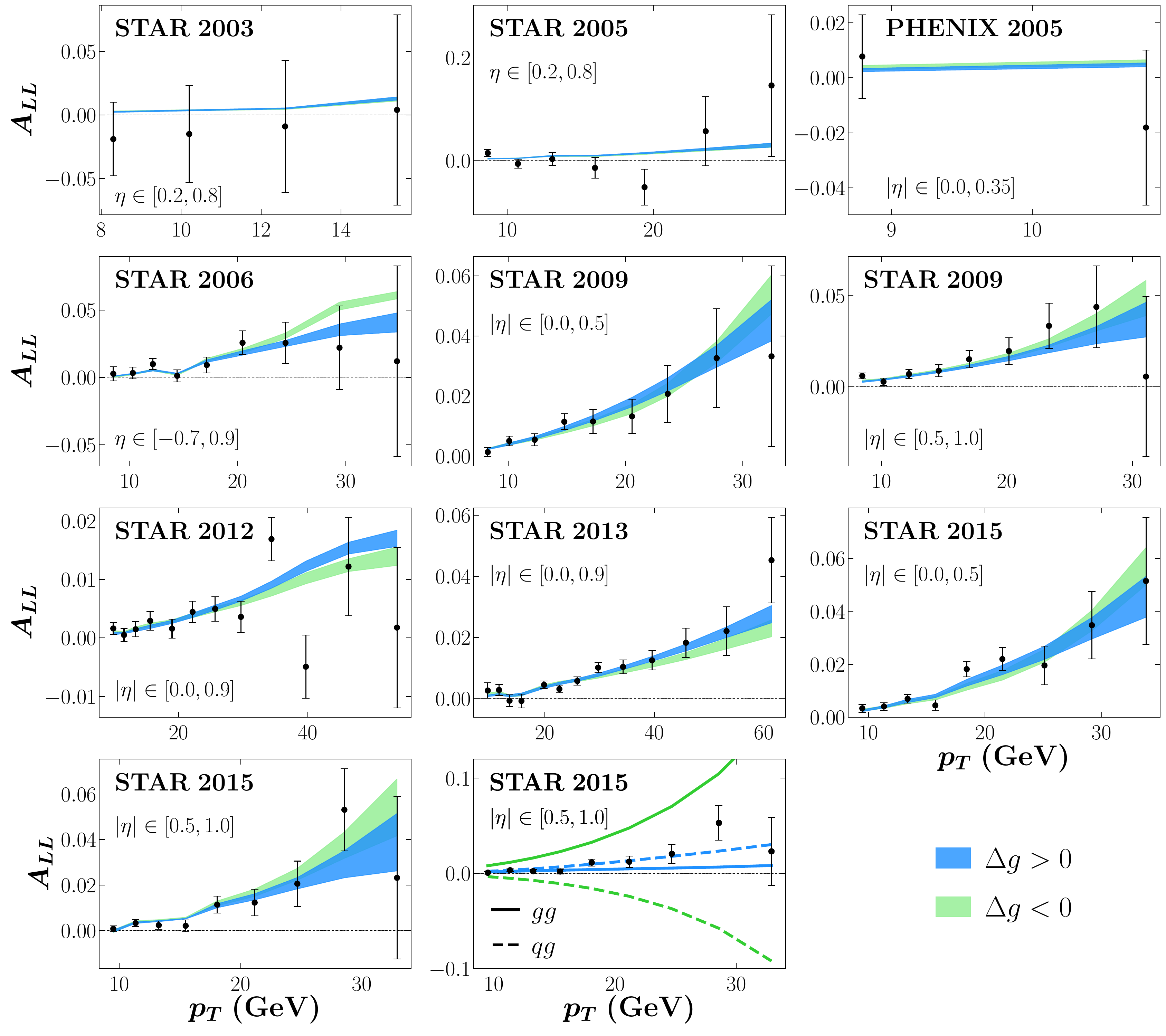}
\caption
[Fit to polarized jet observables from STAR and PHENIX]
{Double longitudinal spin asymmetries $A_{LL}$ in polarized $pp$ collisions from STAR \cite{Abelev:2006uq, Adamczyk:2012qj, Adamczyk:2014ozi, Adam:2019aml, STAR:2021mfd, STAR:2021mqa} and PHENIX \cite{Adare:2010cc} versus jet transverse momentum $p_T$ for bins in pseudorapidity $\eta$.
The data are compared with the JAM global QCD analysis using the SU(3) scenario in Eq.~(\ref{e.su_3}) for the ``positive'' gluon solutions $\Delta g>0$ (blue bands) and ``negative'' gluon solutions $\Delta g<0$ (green bands) with $1\sigma$ uncertainties.
The final panel (with the same data as for the STAR 2015 panel on its left) shows the contributions from the $gg$ (solid lines) and $qg$ channels (dashed lines).
}
\label{f.pjet_fit}
\end{center}
\end{figure}

As for the polarized case, we obtain an excellent description of the spin-dependent observables, including polarized lepton-nucleon DIS and jet production in polarized $pp$ collisions.
For the former, we find a total $\chi^2_{\rm red} \approx 0.9$ (see Ref.~\cite{zhou:thesis} for the corresponding DIS data to theory comparisons).
For the latter, we show in \cref{f.pjet_fit} the inclusive polarization asymmetries, $A_{LL}$, for the STAR and PHENIX data, for the SU(3) scenario, with $\chi_{\rm red}^2 \approx 0.8$ for jet production data in polarized $pp$ collisions.
The two sets of bands represent solutions with $\Delta g > 0$ and $\Delta g < 0$, as we discuss below, with each giving very similar descriptions.
Only the fits to the STAR 2005 \cite{Adamczyk:2012qj} and 2012 \cite{Adam:2019aml} data show noticeable deviations, with $\chi^2_{\rm red} \approx 1.5$ for these sets, which is mostly due to the presence of some outliers in these spectra.

%
The fits to the jet $A_{LL}$ asymmetries are similar for the other scenarios, with $\chi^2_{\rm red}$ values almost identical, as listed in Table~\ref{t.chi_2}.
As may be expected, the less restrictive SU(2) scenario produces moderately wider uncertainty bands at the larger $p_T$ values, $p_T \gtrsim 30$~GeV, due to the relatively larger uncertainties on the helicity PDFs in the absence of the SU(3) flavor symmetry constraint.
Conversely, the more restrictive SU(3)+positivity scenario yields narrower error bands for $p_T \gtrsim 30$~GeV as a result of the significant suppression of the $\Delta g$ solution space from the positivity constraints, as we discuss next.

\begin{figure}[t]
\begin{center}
\includegraphics[width = 0.85 \textwidth]{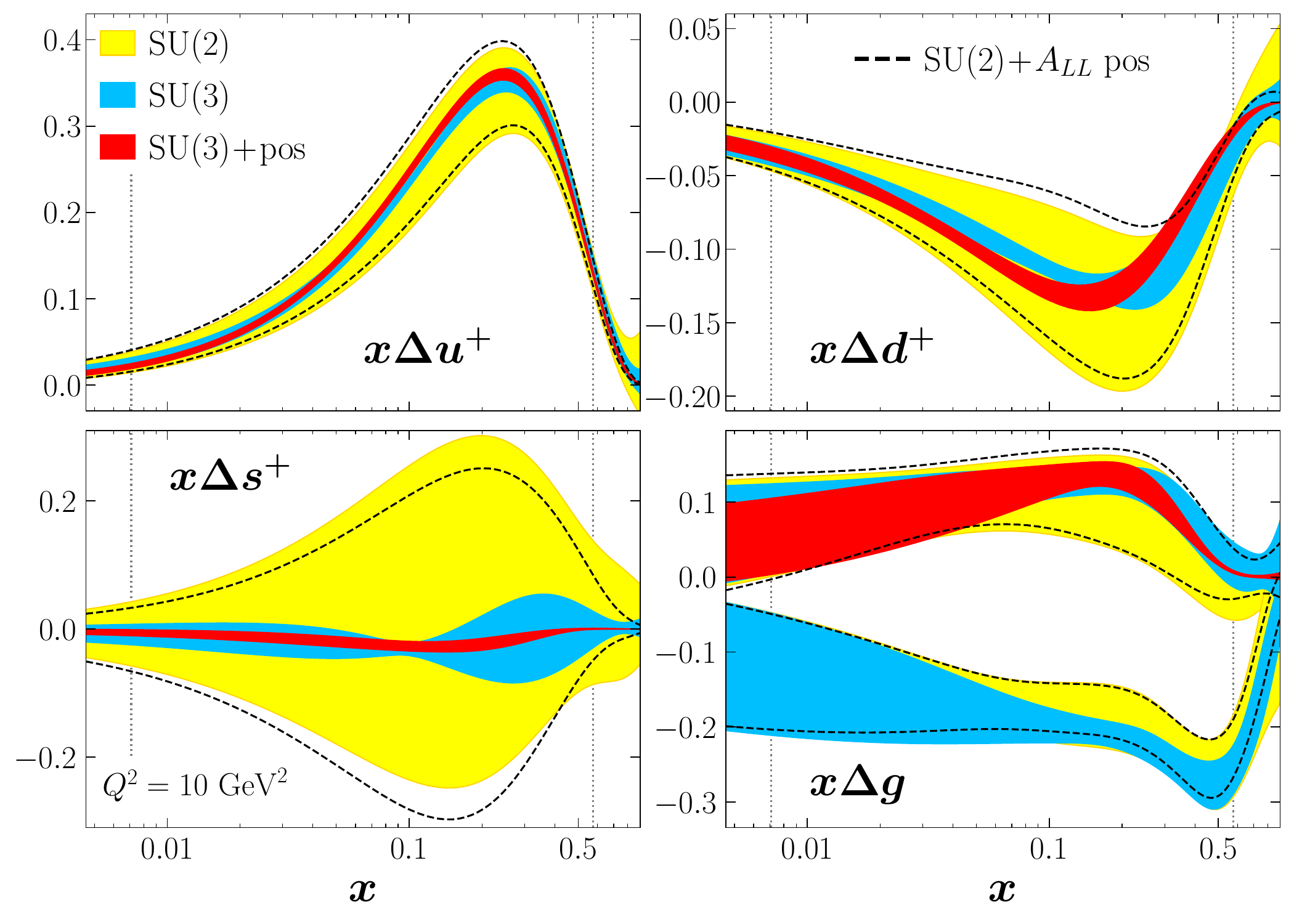}
\caption
[Spin-dependent PDFs fitted with varying theory inputs]
{
Expectations values for spin-dependent $\Delta u^+$, $\Delta d^+$, $\Delta s^+$, and $\Delta g$ PDFs at $Q^2=10$~GeV$^2$ fitted under various theory assumptions according to the SU(2) (yellow 1$\sigma$ bands), SU(3) (blue 1$\sigma$ bands) and SU(3)+positivity (red 1$\sigma$ bands) scenarios, as well as with the SU(2) scenario but filtered to ensure $A_{LL}$ positivity at large $x$ (dashed lines).
The vertical dotted lines indicate the range of parton momentum fractions $x$ constrained by data.}
\label{f.quark_ppdfs}
\end{center}
\end{figure}

To illustrate more explicitly the influence of theoretical assumptions on the PDFs and their uncertainties, we compare in \cref{f.quark_ppdfs} the $\Delta u^+$, $\Delta d^+$, $\Delta s^+$ and $\Delta g$ distributions at $Q^2=10$~GeV$^2$ for the different scenarios with SU(2), SU(3), or SU(3)+positivity constraints.
%
%
For the least constrained fit with only the SU(2) relation in Eq.~(\ref{e.su_2}) imposed, the $\Delta u^+$ and $\Delta d^+$ PDFs are reasonably well determined, while the $\Delta s^+$ distribution has a very large uncertainty and is consistent with zero.
The imposition of the SU(3) relation in Eq.~(\ref{e.su_3}) has a dramatic effect on the polarized quark PDF uncertainties, especially for the $\Delta s^+$ distribution, but also on the nonstrange spin PDFs which have reduced uncertainties.

Imposition of the positivity constraints in Eq.~(\ref{e.positivity}) further reduces the uncertainties on the polarized quark PDFs, especially for the strange quark, and augments somewhat the shape of the $\Delta d^+$ PDF in particular.
The latter effect is induced by assuming a flavor symmetric polarized sea, $\Delta \bar{u} = \Delta \bar{d} = \Delta \bar{s} = \Delta s$, so that changes in $\Delta s$ propagate to the $\Delta u^+$ and $\Delta d^+$ distributions.
Since the absolute values of $\Delta d^+$ are smaller than those of $\Delta u^+$, the impact on the polarized $d$ quark is greater.
The dependence of the strange helicity distribution on theoretical assumptions, such as SU(3) symmetry and positivity, may be reduced with additional experimental data on semi-inclusive DIS and single inclusive $e^+ e^-$ annihilation data, which can provide independent combinations of the quark flavor PDFs~\cite{Ethier:2017zbq, Moffat:2021dji}.

\begin{figure}[t]
\begin{center}
\includegraphics[width = 0.8 \textwidth]{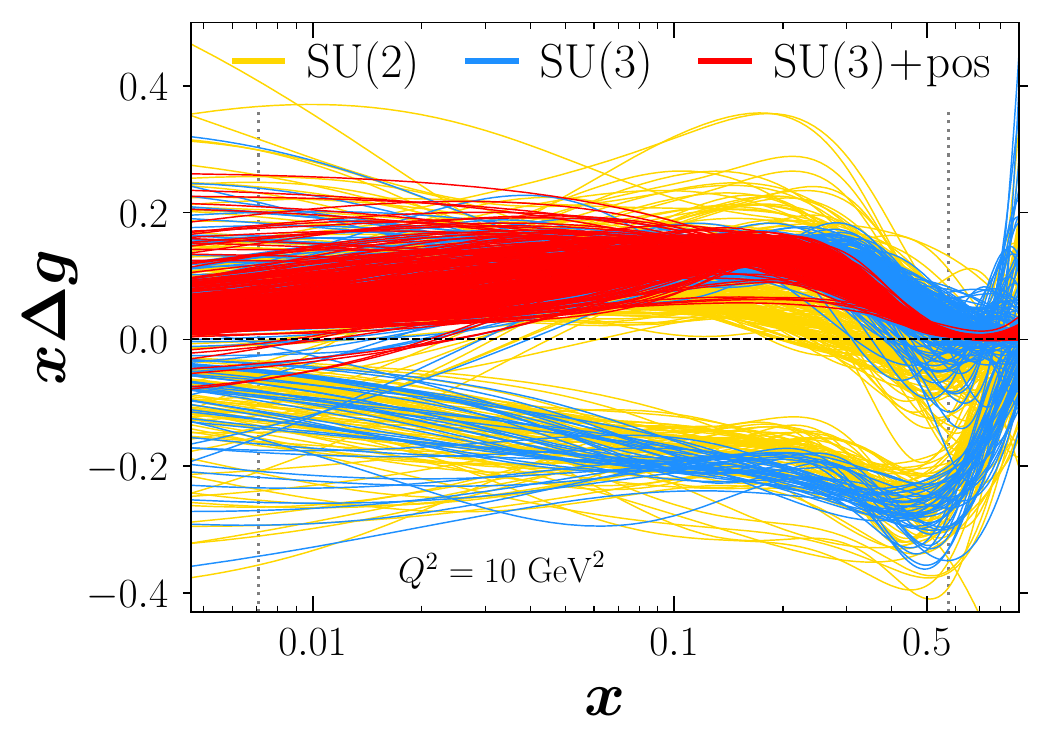}
\caption
[Spin-dependent gluon PDF fitted with varying theory inputs]
{Monte Carlo replicas for the spin-dependent gluon PDF $x \Delta g$ at $Q^2=10$~GeV$^2$ fitted under various theory assumptions according to the SU(2) (yellow lines), SU(3) (blue lines) and SU(3)+positivity (red lines) scenarios, with 300 replicas randomly selected from the total of 723, 647 and 639 for the three scenarios, respectively.
The vertical lines indicate the range of parton momentum fractions $x$ constrained by data.}
\label{f.gluon_ppdf}
\end{center}
\end{figure}

For the gluon helicity distribution, we find in the absence of PDF positivity constraints two distinct sets of solutions that differ in sign, with the positive $\Delta g$ solutions closely resembling results from earlier PDF analyses~\cite{deFlorian:2014yva, Nocera:2014gqa}.
Examining the solution space more closely, we observe that the $\Delta g$ solutions are extremely non-Gaussian for the SU(2) and SU(3) scenarios, as \cref{f.gluon_ppdf} illustrates for the individual replicas.
For the SU(2) scenario, we find 72\% are positive solutions and 28\% are negative solutions, while for the SU(3) case the fractions are 85\% and 15\%, respectively.
With the inclusion of the PDF positivity constraints, when using the negative solutions from the SU(3) scenario as starting points for the minimization process all the replicas remain negative with very large values of {$\chi^2_{\rm red} \approx 5.7$}.
In contrast, all the positive replicas converge again into positive solutions with good {$\chi^2_{\rm red} \approx 1.0$} \cite{zhou:thesis}.
While the optimization process in principle allows migration of negative solutions into positive ones, their starting points are too far away from the positive solutions, and the optimization algorithm does not create sufficient gradient in parameter space to move the negative solutions toward the positive ones.
Despite the lack of convergence of the optimization process with the negative solutions as starting points, from these results we can conclude that the negative solutions are not simultaneously compatible with the data and PDF positivity.

Even though the two $\Delta g$ solutions in the SU(2) and SU(3) scenarios differ dramatically in sign and shape, they are both able to describe the $A_{LL}$ data equally well.
This is illustrated in the STAR 2015 panel of \cref{f.pjet_fit}, where for the $\Delta g > 0$ solution the asymmetry is given by a sum of (small) positive contributions from the $gg$ and $qg$ channels, while for the $\Delta g < 0$ solution the asymmetry results from a cancellation between large positive $gg$ and large negative $qg$ pieces.
As is evident in \cref{f.gluon_ppdf}, our analysis disfavors solutions with $\Delta g = 0$, as well as with small negative $\Delta g$ values, which would generally produce very small $A_{LL}$ values, in contradiction with the data in \cref{f.pjet_fit}.
While the numbers of negative solutions found in the SU(2) and SU(3) scenarios are relatively smaller than the positive ones, their ability to describe well the data indicates that at present the negative solutions cannot be ruled out on phenomenological grounds.


%

In addition to the scenarios discussed above, we also note that some replicas give unphysical values for the polarized DIS asymmetry at kinematics $x \gtrsim 0.8$ and momentum transfer $Q^2 > 50$~GeV$^2$ that are outside the currently measured region, but which could be probed at a future Electron-Ion Collider~\cite{AbdulKhalek:2021gbh}.
After removing these replicas, the result shown in \cref{f.quark_ppdfs} for the SU(2) scenario indicates that the main effect is observed at high $x$ for the quark distributions, while the effect on $\Delta g$ is negligible.
Similarly for the other two scenarios, the impact of imposing the observable positivity on $A_{LL}$ outside measured regions is only marginal.

To further explore the robustness of our findings for $\Delta g$, we also considered the scenario whereby the SU(2) constraint is imposed together with PDF positivity.
In this case the resulting spin-dependent PDFs are found to be very similar to those from the SU(3) + positivity scenario, indicating that the order of imposing the SU(3) flavor symmetry and the positivity constraints does not change the observation of the two types of solutions for $\Delta g$ in Fig.~\ref{f.gluon_ppdf}.

\subsection{Truncated moments}

\begin{figure}[t]
\begin{center}
\includegraphics[width=0.95\textwidth]{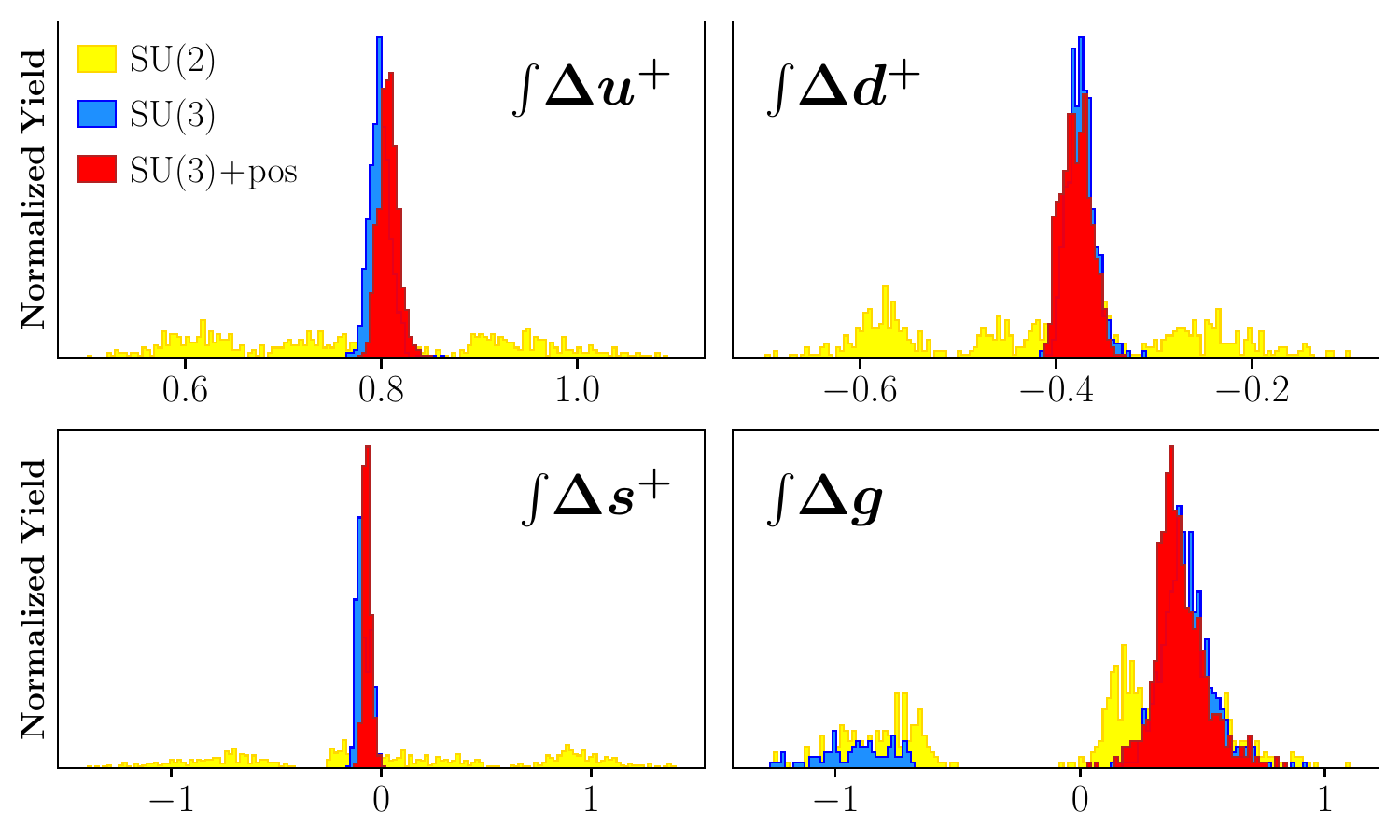}
\caption
[Truncated moments of spin-dependent PDFs with varying theory inputs]
{Distribution of truncated moments of spin-dependent quark and gluon PDFs, integrated from $x_{\min}=0.0071$ to 1, extracted from our global analysis under the different theoretical scenarios of SU(2) (yellow), SU(3) (blue) and SU(3)+positivity (red histograms) at a scale $Q^2 = 10$~GeV$^2$.}
\label{f.ppdf_moments}
\end{center}
\end{figure}

Along with visualizing the $x$ dependence of the spin-dependent PDFs, a complementary way to assess the impact of the theoretical inputs on the quark and gluon helicities is to consider the truncated moments of the PDFs, defined as
\begin{eqnarray}
\label{e.moments}
\int \Delta q^+ &\equiv& \int_{x_{\rm min}}^1 \dd{x} \Delta q^+ ,
\qquad
\int \Delta g\,  \equiv\, \int_{x_{\rm min}}^1 \dd{x} \Delta g ,
\end{eqnarray}
where $x_{\min}$ is the lower limit of the integral.
We choose the lower limit to be the smallest $x$ value to which polarized data have sensitivity, $x_{\min} = 0.0071$ (see \cref{f.kin}).
In \cref{f.ppdf_moments} we show the distribution of the quark and gluon truncated moments at a scale $Q^2 = 10$~GeV$^2$.
The distributions of the quark truncated moments for the SU(2) scenario are rather broad (the SU(2) nonsinglet $\int\Delta u^+ - \int\Delta d^+$ is much more constrained though), and the gluon truncated moment displays the clear gap between the positive and negative solutions seen in \cref{f.gluon_ppdf}.
The reduction of the uncertainties with the imposition of SU(3) symmetry is quite striking for the quark moments, but does not qualitatively alter the distribution of gluon truncated moments, other than a slightly stronger peak in the positive $\Delta g$ solution space.
With the positivity constraints imposed, on the other hand, the negative $\Delta g$ solution is eliminated, with a prominent single peak around $\int \Delta g \approx 0.4$.

The central values and uncertainties of the truncated moments of all the quark and gluon flavors are shown in \cref{t.moments} for the three different scenarios.
For the gluon in the SU(2) and SU(3) constraints scenarios we show the individual contributions to the truncated moment from the positive and negative $\Delta g$ solutions, along with the result of combining the two sets of solution.
While the central values of the polarized quark moments do not vary much across the scenarios, the values of the gluon moments depend strongly on the theoretical assumptions.
For the positive $\Delta g$ solutions, the truncated moments remain at $\sim 0.4$, but together with the negative $\Delta g$ solutions for the SU(2) and SU(3) scenarios, which yield $\approx -0.9$, the combined moment ranges from $\approx 0$ to 0.25, with large uncertainty $\approx 0.5$.

\begin{table}[b]
\vspace*{0.5cm}
\caption
[Truncated moment $\int \Delta g$ integrated from 0.01]
{
Truncated moments and uncertainties of the quark and gluon PDFs integrated from $x_{\min}=0.0071$ to 1 at $Q^2=10$~GeV$^2$ for the SU(2), SU(3), and SU(3)+positivity scenarios. For the SU(2) and SU(3) cases, the individual positive and negative $\int\Delta g$ contributions are also shown separately. \\
}
\begin{tabular}{c|rr|rr|r}
\hline
~~$\bm{\int\!\Delta f}$~~
& \multicolumn{2}{c|}{\footnotesize{\bf SU(2)}}
& \multicolumn{2}{c|}{\footnotesize{\bf SU(3)}}
& ~~~~\footnotesize{\bf SU(3)+pos}~~~~ \\
\hline
$\Delta u^+$ 
& \multicolumn{2}{r|}{0.8(1)~~~~~~~~~~}
& \multicolumn{2}{r|}{0.80(1)~~~~~~~~~}
& 0.81(1)~~~~~~
\\
$\Delta d^+$ 
& \multicolumn{2}{r|}{$-0.4(1)$~~~~~~~~~~}
& \multicolumn{2}{r|}{$-0.37(1)$~~~~~~~~~}
& $-0.38(2)$~~~~~~
\\ 
$\Delta s^+$ 
& \multicolumn{2}{r|}{0.1(7)~~~~~~~~~~}
& \multicolumn{2}{r|}{$-0.08(3)$~~~~~~~~~} 
& $-0.07(2)$~~~~~~
\\ 
\hline
\multirow{3}{*}{$\Delta g$~~} 
& \multicolumn{2}{r|}{0.0(6)~~~~~~~~~~}  
& \multicolumn{2}{r|}{0.3(5)\,~~~~~~~~~~} 
& \multirow{3}{*}{0.39(9)}~~~~~~
\\ 
\cline{2-5}
& ~~$\Delta g>0$~ 
& ~$\Delta g<0$~~
& ~~$\Delta g>0$~ 
& ~$\Delta g<0$~~ 
&
\\
& ~0.4(2)~ 
& $-0.8(2)$~~
& ~0.4(1)~ 
& $-0.9(2)$~~
&
\\
\hline
\end{tabular}
\label{t.moments}
\end{table}

Compared with the results from the DSSV14 analysis~\cite{deFlorian:2014yva}, which gave $\int \Delta g = 0.20(5)$ for $x_{\min}=0.05$, our positive $\Delta g$ solutions give values comparable to the DSSV14 result for all scenarios, ranging from 0.20(13) for the least restrictive SU(2) case to 0.25(3) for the most restrictive SU(3)+positivity case.
Combining positive and negative $\Delta g$ solutions, however, the gluon truncated moments over this range are $0.0(4)$ and 0.1(3) for the SU(2) and SU(3) scenarios, in clear contrast to the DSSV14 result.
This strong dependence on the theoretical assumptions used in the analysis suggests that additional data with greater sensitivity to the shape and sign of $\Delta g$ are needed before definitive, experiment-driven conclusions about gluon polarization can be reached.


\subsection{Helicity-basis PDFs}
\label{sec.helicity-basis}

\begin{figure}[t]
\begin{center}
\includegraphics[width = 0.99 \textwidth]{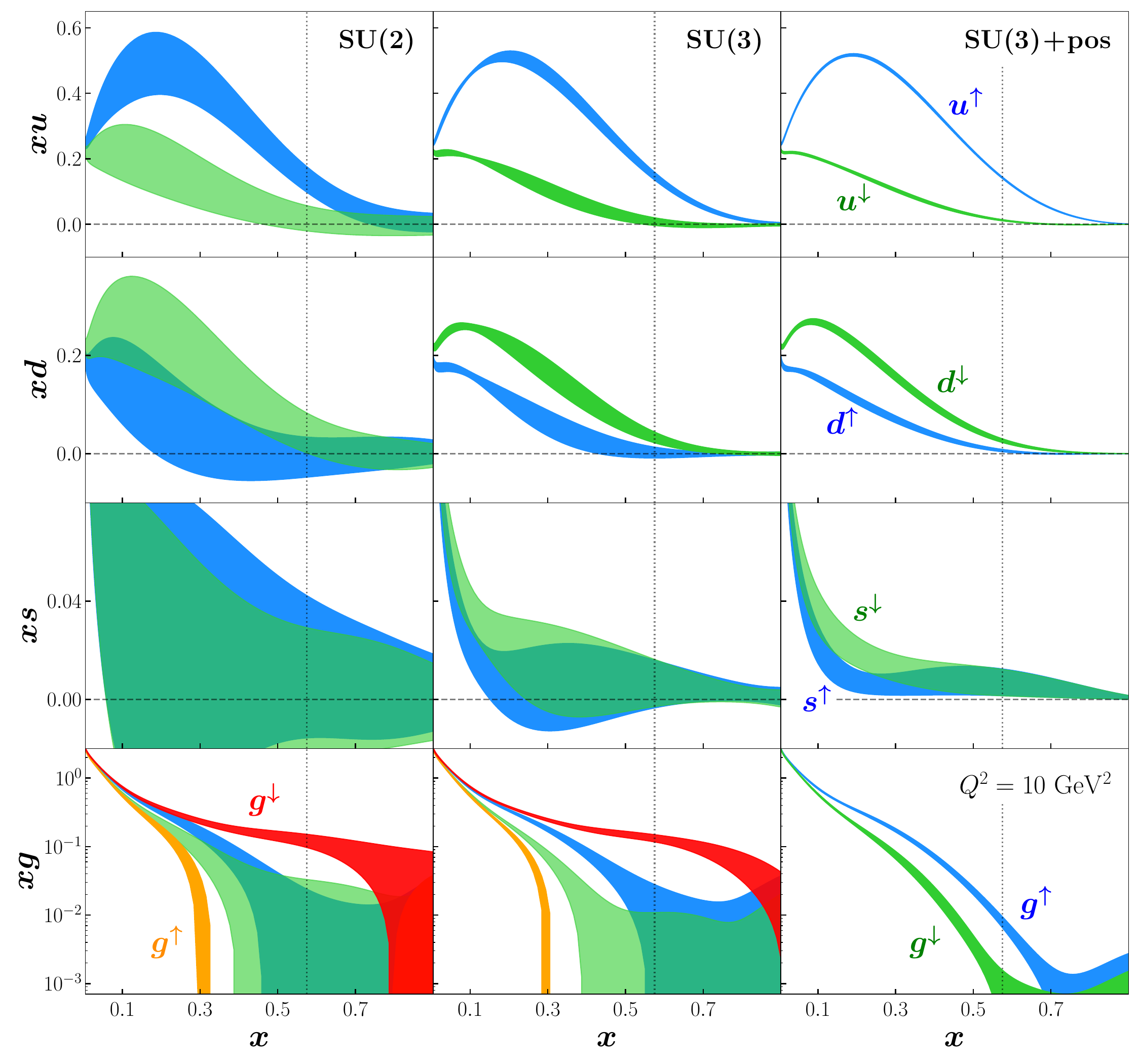}
\caption
[helicity basis from simultaneous fits]
{Helicity-basis PDFs for helicity-aligned $f^{\uparrow}$ (blue bands) and antialigned $f^{\downarrow}$ (green bands) distributions for $f=u$, $d$, $s$ and $g$ at $Q^2 = 10$~GeV$^2$, in the SU(2), SU(3), and SU(3)+positivity scenarios.
For the gluon helicity basis distributions, the positive and negative solutions for \mbox{\{$g^{\uparrow}$, $g^{\downarrow}$\}} are indicated by the \{blue, green\} and \{red, orange\} bands, respectively.
The vertical dotted lines indicate the maximum $x$ constrained by polarized data. Note also the logarithmic scale for the gluon along the ordinate.}
\label{f.helicity}
\end{center}
\end{figure}

Having obtained good agreement with both the unpolarized and polarized world datasets (see \cref{t.chi_2}), we can now analyze the simultaneously extracted spin-averaged and spin-dependent PDFs from the combined analysis.
Using Eqs.~(\ref{e.helicitybasis}), the distributions with spins parallel ($f^\uparrow$) and antiparallel ($f^\downarrow$) to the proton spin can be extracted for the first time with a consistent treatment of PDF uncertainties.
In Fig.~\ref{f.helicity} we show the helicity-basis PDFs for all the $u$, $d$, $s$ and $g$ flavors at $Q^2 = 10$~GeV$^2$ in the SU(2), SU(3), and SU(3)+positivity scenarios.

For the SU(2) scenario, the $u^{\uparrow}$ and $u^{\downarrow}$ distributions, and to some degree the $d^{\uparrow}$ and $d^{\downarrow}$ distributions, are reasonably distinguishable, while the strange and gluon helicity-basis distributions for the most part cannot be distinguished. 
With the imposition of additional constraints, the uncertainties decrease and the helicity-basis PDFs $f^{\uparrow}$ and $f^{\downarrow}$ for each flavor become more clearly separated.
Imposing the SU(3) constraint reduces the uncertainties of the helicity-basis PDFs for all the light quark flavors as a result of the extra constraint provided by $a_8$ in \cref{e.su_3}.
Moreover, the addition of positivity constraints restricts each helicity-basis PDF to remain positive, and in the process suppresses uncertainties in the high-$x$ region.

For the gluon helicity-basis distributions, Fig.~\ref{f.helicity} shows both the positive and negative $\Delta g$ sets of solutions separately for the SU(2) and SU(3) scenarios.
The $g^{\uparrow}$ and $g^{\downarrow}$ PDFs can be clearly distinguished from each other for both solutions, although it would not be distinguished as well if these were combined.
For the SU(3)+positivity scenario, since the negative solutions are no longer present, the $g^{\uparrow}$ and $g^{\downarrow}$ distributions are clearly separated.

The degree to which the different helicity-basis PDFs in \cref{f.helicity} can be delineated can be more accurately quantified by considering an ``area under the curve'' (AUC) plot.
The AUC is defined as the area under a receiver operating characteristic (ROC) curve~\cite{Egan75, Fawcett06} (see Appendix~\ref{s.auc} for details), and is often used to visualize the discrimination power in binary classification problems.
The closer the AUC approaches 1, the better the discrimination between the $f^{\uparrow}$ and $f^{\downarrow}$ distributions, and, conversely, the closer the AUC approaches 1/2, the more difficult it is to discriminate.

\begin{figure}[t]
\begin{center}
\includegraphics[width = 0.95 \textwidth]{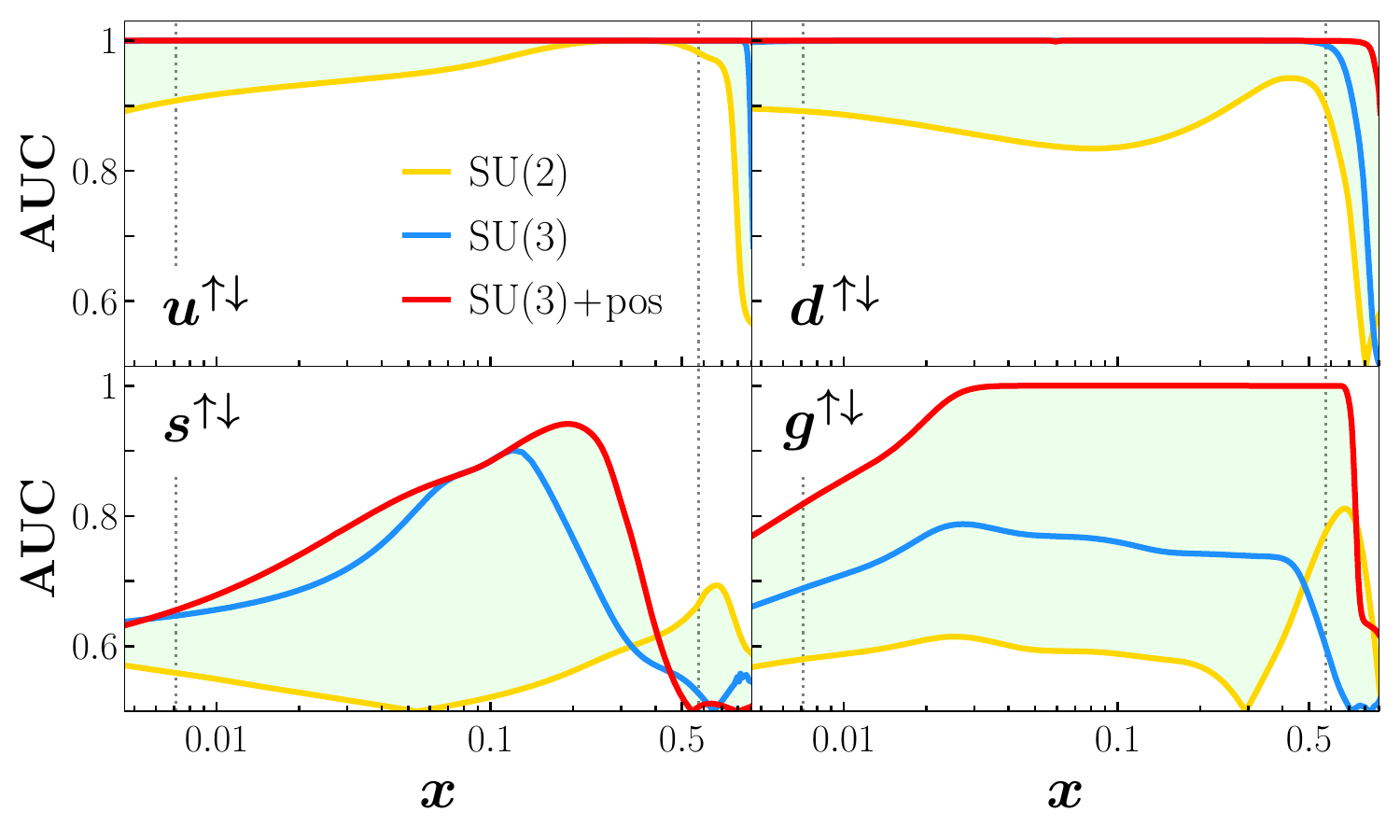}
\caption
[AUC plot for the helicity basis]
{The area under the curve (AUC) of the receiver operating characteristic (ROC) for the helicity-basis PDFs $f^\uparrow$ and $f^\downarrow$ for $f=u$, $d$, $s$, and $g$ at $Q^2 = 10$~GeV$^2$, for the SU(2) (yellow lines), SU(3) (blue lines), and SU(3)+positivity (red lines) constraint scenarios. The discrimination between $f^\uparrow$ and~$f^\downarrow$ improves as AUC $\to 1$. For clarity the region between the lines has been shaded (light green).
The vertical dotted lines indicate the range of parton momentum fractions constrained by data.}
\label{f.auc}
\end{center}
\end{figure}

In Fig.~\ref{f.auc} we show the AUC plot for the $f^\uparrow$ and $f^\downarrow$ helicity-basis PDFs for $f=u$, $d$, $s$, and $g$ at $Q^2 = 10$~GeV$^2$, for the SU(2), SU(3), and SU(3)+positivity scenarios. 
Firstly, we note that the discrimination between the $u^\uparrow$ and $u^\downarrow$ PDFs is not affected significantly by the theory inputs, which reflects that both the spin-averaged and spin-dependent $u$-quark distributions are already well constrained empirically.
The discrimination between $d^\uparrow$ and $d^\downarrow$, on the other hand, receives discernible improvement from the SU(3) flavor symmetry constraint, which produces a reduction of the uncertainty in the polarized $\Delta d$ PDF (see \cref{f.quark_ppdfs}).
For the $s$-quark helicity-basis PDFs, imposing SU(3) symmetry makes a large improvement to their discrimination in the data-sensitive region $0.01 \lesssim x \lesssim 0.5$, driven by the significant reduction of the uncertainty for the $\Delta s^+$ PDF.

The positivity constraints make almost no improvement to the discrimination between the $u^\uparrow$ and $u^\downarrow$ or $d^\uparrow$ and $d^\downarrow$ distributions, given that the helicity-basis PDFs of both flavors are already  well separated, and the positivity constraints only reduce their uncertainties.
For the $s$-quark helicity PDFs, on the other hand, one may have expected that, given the significant reduction of the $\Delta s^+$ uncertainty and the less well discriminated $s^\uparrow$ and $s^\downarrow$ PDFs in the SU(3) scenario, the positivity constraints should discriminate between $s^\uparrow$ and $s^\downarrow$ more effectively.
However, because the spin-averaged strange distributions are less well constrained at high~$x$, the reduction of the $\Delta s^+$ uncertainty does not result in better discrimination between $s^\uparrow$ and $s^\downarrow$.

While the effects of the theory assumptions on the gluon distributions have been discussed extensively in the previous sections, we can obtain further information on their impact through the AUC representation in \cref{f.auc}.
Starting with the SU(2) scenario, without additional constraints the $g^\uparrow$ and $g^\downarrow$ distributions are barely distinguishable.
After imposing the SU(3) constraint the reduction of the $\Delta g$ uncertainties (see \cref{f.gluon_ppdf}) discernibly improves the discrimination power.
Finally, with the positivity constraints added, the negative $\Delta g$ solution is eliminated, and the individual helicity-basis PDFs can be clearly distinguished in the experimentally measured region, $0.01 \lesssim x \lesssim 0.5$.
This further illustrates our main conclusion, which is that theoretical inputs, especially SU(3) flavor symmetry and PDF positivity, can introduce significant bias into the extraction of PDFs that are not well constrained by experimental data, such as the $\Delta s$ and $\Delta g$ spin distributions.

\section{Conclusion}
\label{s.conclusion}

We have performed the first simultaneous global QCD analysis of spin-averaged and spin-dependent PDFs in the nucleon within the JAM multi-step Monte Carlo framework, focusing in particular on the extraction of the gluon polarization.
Good fits to the world unpolarized and polarized DIS data, as well as Drell-Yan data and inclusive jet production cross sections and asymmetries in hadronic collisions, were achieved, with a global reduced $\chi^2_{\rm red} \approx 1.1$.
Our study was the first time that unpolarized RHIC $pp$ jet cross sections were included together with the polarized $pp$ jet data, along with the previous jet measurements in $p\bar p$ collisions at the Tevatron.
While the direct impact of the unpolarized data on the spin-dependent PDFs is not significant, a simultaneous description of both observables is needed in order to delineate the kinematic domain of applicability of the collinear factorization framework.

Our study critically assessed the impact of theoretical assumptions on the determination of the gluon polarization $\Delta g$, including scenarios in which SU(2) or SU(3) flavor symmetry is assumed for axial charges determined from neutron and hyperon beta-decays, as well as the imposition of positivity on PDFs, which has been debated recently in the literature.
The least biased scenario involving only the SU(2) constraint produced relatively large PDF uncertainties, especially for the polarized strange and gluon distributions.
In particular, we found two distinct types of solutions for the gluon polarization, of opposite sign, each giving almost identical descriptions of experimental spin-dependent observables, including the double polarization asymmetry $A_{LL}$ in inclusive jet production.

With the SU(3) flavor symmetry constraint, the uncertainties on $\Delta s$ were reduced significantly, although the $\Delta g$ distribution was largely unaffected.
It was only with the further addition of PDF positivity that the negative $\Delta g$ solution could be eliminated and results resembling those found in earlier literature recovered.
We identify this as a bias introduced in the extraction of spin-dependent PDFs, and in the absence of a clear theoretical requirement for PDFs to be positive at all values of $x$, we conclude that data-driven analysis alone does not constrain the gluon polarization to be uniquely positive.

Such a conclusion naturally has profound implications for our understanding of the proton spin decomposition, and provides greater urgency to identifying other means to constrain the gluon helicity, as well as the quark and gluon orbital angular momentum.
Further data are needed to obtain clarity on this issue, including new observables that are linearly sensitive to gluon helicity distributions.
An example of this may be polarized semi-inclusive DIS with the production of large transverse momentum hadrons, in which the gluon polarization enters at the same order as the polarized quark contribution~\cite{Whitehill22}.
Alternatively, inclusive charm meson production in polarized DIS could provide another path toward constraining the polarized glue through the photon-gluon fusion process~\cite{COMPASS:2012mpe, Anderle:2021hpa}.

A novel new feature of our simultaneous analysis of polarized and unpolarized observables is the ability to systematically extract the individual helicity-aligned and antialigned PDFs with a consistent treatment of uncertainties.
Our demonstration of the current empirical situation using the AUC plot illustrates the impact of various theoretical assumptions on our ability to discriminate between different helicity-basis PDFs.
It confirms that, while imposing SU(3) flavor symmetry improves the discrimination between the $s^\uparrow$ and $s^\downarrow$ PDFs, the ability to separate clearly the $g^\uparrow$ and $g^\downarrow$ distributions requires additional assumptions about PDF positivity.
We anticipate that future high-precision data from existing and planned facilities, including the Electron-Ion Collider~\cite{AbdulKhalek:2021gbh}, will elucidate the question of the gluon polarization and the proton spin decomposition more definitively. \\ \\

\section*{Acknowledgements}

We thank C.~Andres, P.~Barry, C.~Cocuzza, J.~Collins, J.~Owens, T.~Rogers and F.~Ringer for helpful comments and discussions, and W.~Vogelsang for providing the code for the calculation of the jet observables.
This work was supported by the US Department of Energy (DOE) contract No.~DE-AC05-06OR23177, under which Jefferson Science Associates, LLC operates Jefferson Lab. 
The work of N.S. was supported by the DOE, Office of Science, Office of Nuclear Physics in the Early Career Program.
The work of Y.Z. is also supported by the Guangdong Major Project of Basic and Applied Basic Research No.~2020B0301030008, the National Natural Science Foundation of China under Grant No.~12022512, No.~12035007.
The majority of Y.Z.'s work was done while he was a graduate student at William \& Mary and Jefferson Lab.

\clearpage
\appendix
\section{Area Under the ROC Curve}
\label{s.auc}

The receiver operating characteristic (ROC) and the corresponding area under the ROC curve (AUC) are often used in visualizing the discrimination in binary classification problems~\cite{Egan75, Fawcett06}.
In this appendix we will use a simple example to illustrate the application of ROC and AUC plots for PDF discrimination.

\begin{figure}[t]
\begin{center}
\includegraphics[width = 0.85 \textwidth]{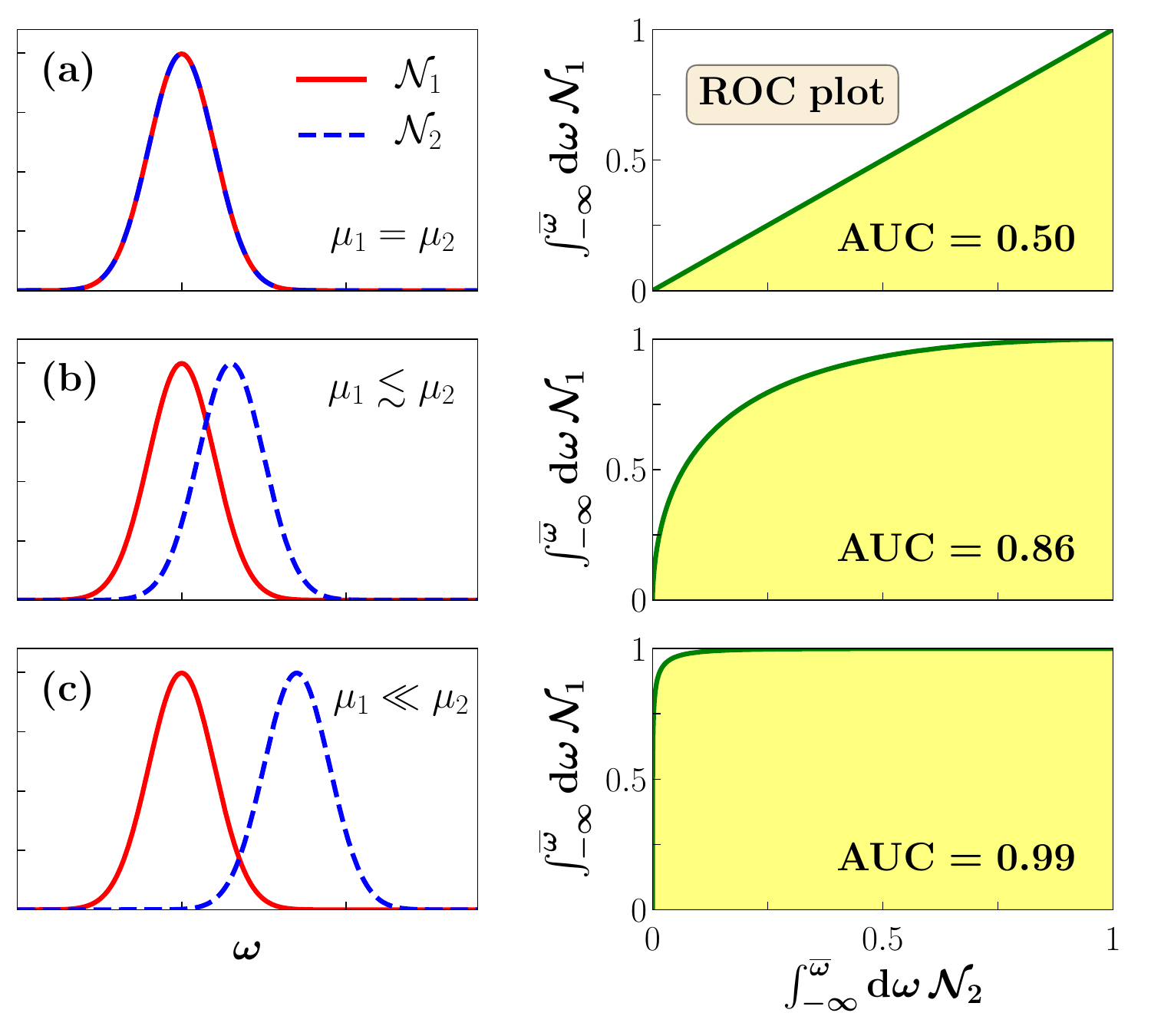}
\caption
[Demonstration of ROC and AUC]
{
Illustration of ROC and AUC plots for Gaussian distributions $\mathcal{N}_1$ and $\mathcal{N}_2$ (red solid and blue dashed lines) to be discriminated from each other (left panels), and the corresponding ROC curves (green lines) and AUC values (right panels) for different mean values $\mu_1$ and $\mu_2$:
{\bf (a)}~$\mu_1 = \mu_2$,
{\bf (b)}~$\mu_1 \lesssim \mu_2$,
{\bf (c)}~$\mu_1 \ll \mu_2$.
}
\label{f.roc}
\end{center}
\end{figure}

Defining two normalized Gaussian distributions $\mathcal{N}_1$ and $\mathcal{N}_2$ in a common variable $\omega$, with central values $\mu_1$ and $\mu_2$ and widths $\sigma_1 = \sigma_2$, in \cref{f.roc} we examine how well these can be discriminated from each other for varying $\mu_1-\mu_2$.
We consider the cumulative integration values of $\mathcal{N}_2$ versus those of $\mathcal{N}_1$, defined as
\begin{equation}
\int_{-\infty}^{\overline\omega} \dd{\omega}\, \mathcal{N}_i(\omega)
= \int_{-\infty}^{\overline\omega} \frac{\dd{\omega}}{\sqrt{2 \pi} \sigma_i} \exp \bigg(\!-\frac{(\omega - \mu_i)^2}{2 \sigma_i^2} \bigg)
\quad {\rm for\ }\ \overline\omega \in \big(- \infty , \infty \big),
\end{equation}
for $i=1, 2$.
When the two distributions $\mathcal{N}_1$ and $\mathcal{N}_2$ overlap entirely with each other, $\mu_1=\mu_2$, their cumulative integration values are of course identical, and these produce a diagonal line in the ROC plot of $\int_{- \infty}^{\overline{\omega}} \dd{\omega} \, \mathcal{N}_2$ versus $\int _{- \infty}^{\overline{\omega}} \dd{\omega} \, \mathcal{N}_1$ in \cref{f.roc} (top row).
In this case the AUC value is exactly 1/2.

When the mean values of $\mathcal{N}_1$ and $\mathcal{N}_2$ differ, but not dramatically, $\mu_1 \lesssim \mu_2$ (middle row), the distributions start to deviate from the each other, and the ROC bends away from the diagonal since $\int_{-\infty}^{\overline{\omega}} \dd{\omega}\, \mathcal{N}_2$ reaches its maximum value slower than $\int_{-\infty}^{\overline{\omega}} \dd{\omega} \, \mathcal{N}_1$.
The AUC value here increases away from the minimum of 1/2.

Finally, when the mean values are clearly separated from each other, $\mu_1 \ll \mu_2$ (bottom row of \cref{f.roc}), the ROC deviates significantly from the diagonal and the AUC value approaches~1.
Of course, if one had $\mu_1 \gtrsim \mu_2$ or $\mu_1 \gg \mu_2$, the ROC would be curved downwards and the AUC would deviate from 1/2 while approaching 0; in this case the $1-{\rm AUC}$ would usually be used in order to keep the figures intuitive.

For the binary classification problem in Sec.~\ref{sec.helicity-basis}, the goal is to evaluate the degree to which the helicity-basis PDFs $f^\uparrow$ and $f^\downarrow$ can be discriminated from each other at every point in the parton momentum fraction $x$.
The AUC plot in \cref{f.auc} then represents the quality of discrimination as a function of $x$ for the different theoretical scenarios assuming SU(2) symmetry, SU(3) symmetry, or SU(3) + PDF positivity.

\clearpage


\begin{thebibliography}{99}

\bibitem{EuropeanMuon:1987isl}
J.~Ashman \textit{et al.}, 
\href{https://doi.org/10.1016/0370-2693(88)91523-7}
{Phys. Lett. B \textbf{206}, 364 (1988)}.

\bibitem{Altarelli:1988nr}
G.~Altarelli and G.~G.~Ross,
\href{https://doi.org/10.1016/0370-2693(88)91335-4}
{Phys. Lett. B \textbf{212}, 391 (1988)}.

\bibitem{Carlitz:1988ab}
R.~D.~Carlitz, J.~C.~Collins and A.~H.~Mueller,
\href{https://doi.org/10.1016/0370-2693(88)91474-8}
{Phys. Lett. B \textbf{214}, 229 (1988)}.

\bibitem{Ellis:1973kp}
J.~R.~Ellis and R.~L.~Jaffe,
\href{https://doi.org/10.1103/PhysRevD.9.1444}
{Phys. Rev. D \textbf{9}, 1444 (1974)}
[Erratum: \href{https://doi.org/10.1103/PhysRevD.10.1669.4}
{Phys. Rev. D \textbf{10}, 1669 (1974).}

\bibitem{Close:1993mv}
F.~E.~Close and R.~G.~Roberts,
\href{https://doi.org/10.1016/0370-2693(93)90673-6}
{Phys. Lett. B \textbf{316}, 165 (1993).}

\bibitem{Anselmino:1994gn}
M.~Anselmino, A.~Efremov and E.~Leader,
\href{https://doi.org/10.1016/0370-1573(95)00011-5}
{Phys. Rep. \textbf{261}, 1 (1995)}.
[Erratum: \href{https://doi.org/10.1016/S0370-1573(96)00027-0}
{Phys. Rep. \textbf{281}, 399 (1997)}].

\bibitem{Lampe:1998eu}
B.~Lampe and E.~Reya,
\href{https://doi.org/10.1016/S0370-1573(99)00100-3}
{Phys. Rep. \textbf{332}, 1 (2000)}.

\bibitem{Aidala:2012mv}
C.~A.~Aidala, S.~D.~Bass, D.~Hasch and G.~K.~Mallot,
\href{https://doi.org/10.1103/RevModPhys.85.655}
{Rev. Mod. Phys. \textbf{85}, 655 (2013)}.

\bibitem{Ethier:2017zbq}
J.~J.~Ethier, N.~Sato and W.~Melnitchouk,
\href{http://doi.org/10.1103/PhysRevLett.119.132001}
{Phys. Rev. Lett. \textbf{119}, 132001 (2017)}.

\bibitem{Diehl:2003ny}
M.~Diehl,
\href{https://doi.org/10.1016/j.physrep.2003.08.002}
{Phys. Rep. \textbf{388}, 41 (2003)}.

\bibitem{Leader:2013jra}
E.~Leader and C.~Lorc\'e,
\href{https://doi.org/10.1016/j.physrep.2014.02.010}
{Phys. Rep. \textbf{541}, 163 (2014)}.

\bibitem{Muller:1994ses}
D.~M\"uller, D.~Robaschik, B.~Geyer, F.~M.~Dittes and J.~Ho\v{r}ej\v{s}i,
\href{https://doi.org/10.1002/prop.2190420202}
{Fortsch. Phys. \textbf{42}, 101 (1994)}.

\bibitem{Ji:1996nm}
X.~Ji,
\href{https://doi.org/10.1103/PhysRevD.55.7114}
{Phys. Rev. D \textbf{55}, 7114 (1997)}.

\bibitem{Radyushkin:1996nd}
A.~V.~Radyushkin,
\href{https://doi.org/10.1016/0370-2693(96)00528-X}
{Phys. Lett. B \textbf{380}, 417 (1996)}.

\bibitem{Alexandrou:2017oeh}
C.~Alexandrou {\it et al.},
\href{https://doi.org/10.1103/PhysRevLett.119.142002}
{Phys. Rev. Lett. \textbf{119}, 142002 (2017)}.

\bibitem{Kumericki:2016ehc}
K.~Kumericki, S.~Liuti and H.~Moutarde,
\href{https://doi.org/10.1140/epja/i2016-16157-3}
{Eur. Phys. J. A \textbf{52}, 157 (2016)}.

\bibitem{Abelev:2006uq}
B.~I.~Abelev \textit{et al.},
\href{https://doi.org/10.1103/PhysRevLett.97.252001}
{Phys. Rev. Lett. \textbf{97}, 252001 (2006)}.

\bibitem{Abelev:2007vt}
B.~I.~Abelev \textit{et al.},
\href{https://doi.org/10.1103/PhysRevLett.100.232003}
{Phys. Rev. Lett. \textbf{100}, 232003 (2008)}.

\bibitem{Adamczyk:2012qj}
L.~Adamczyk \textit{et al.},
\href{https://doi.org/10.1103/PhysRevD.86.032006}
{Phys. Rev. D \textbf{86}, 032006 (2012)}.

\bibitem{Adamczyk:2014ozi}
L.~Adamczyk \textit{et al.},
\href{https://doi.org/10.1103/PhysRevLett.115.092002}
{Phys. Rev. Lett. \textbf{115}, 092002 (2015)}.

\bibitem{Adam:2019aml}
J.~Adam \textit{et al.},
\href{https://doi.org/10.1103/PhysRevD.100.052005}
{Phys. Rev. D \textbf{100}, 052005 (2019)}.

\bibitem{STAR:2021mfd}
M.~Abdallah \textit{et al.},
\href{https://doi.org/10.1103/PhysRevD.103.L091103}
{Phys. Rev. D \textbf{103}, L091103 (2021)}.

\bibitem{STAR:2021mqa}
M.~S.~Abdallah \textit{et al.}, 
\href{http://arXiv.org/abs/2110.11020}
{arXiv:2110.11020 [hep-ex]}.

\bibitem{Adare:2010cc}
A.~Adare \textit{et al.},
\href{https://doi.org/10.1103/PhysRevD.84.012006}
{Phys. Rev. D \textbf{84}, 012006 (2011)}.

\bibitem{Jager:2004jh}
B.~J\"{a}ger \textit{et al.},
\href{https://doi.org/10.1103/PhysRevD.70.034010}
{Phys. Rev. D \textbf{70}, 034010 (2004)}.

\bibitem{deFlorian:2014yva}
D.~de Florian, R.~Sassot, M.~Stratmann and W.~Vogelsang,
\href{http://doi.org/10.1103/PhysRevLett.113.012001}
{Phys. Rev. Lett. \textbf{113}, 012001 (2014)}.

\bibitem{Sato:2019yez}
N.~Sato, C.~Andres, J.~J.~Ethier, and W.~Melnitchouk,
\href{http://doi.org/10.1103/PhysRevD.101.074020}
{Phys. Rev. D \textbf{101}, 074020 (2020)}.

\bibitem{Moffat:2021dji}
E.~Moffat, W.~Melnitchouk, T.~C.~Rogers and N.~Sato,
\href{http://doi.org/10.1103/PhysRevD.104.016015}
{Phys. Rev. D \textbf{104}, 016015 (2021)}.

\bibitem{Cocuzza:2021}
C.~Cocuzza, W.~Melnitchouk, A.~Metz and N.~Sato,
in preparation (2021).

\bibitem{Nocera:2014gqa}
E.~R.~Nocera \textit{et al.}, 
\href{https://doi.org/10.1016/j.nuclphysb.2014.08.008}
{Nucl. Phys. \textbf{B887}, 276 (2014)}.

\bibitem{Candido:2020yat}
A.~Candido, S.~Forte and F.~Hekhorn,
\href{https://doi.org/10.1007/JHEP11(2020)129}
{J. High Energy Phys. 11 (2020) 129}.

\bibitem{Collins:2021vke}
J.~C.~Collins, T.~C.~Rogers and N.~Sato,
\href{http://arXiv.org/abs/2111.01170}
{arXiv:2111.01170 [hep-ph]}.

\bibitem{Adolph:2015saz}
C.~Adolph \textit{et al.},
\href{https://doi.org/10.1016/j.physletb.2015.11.064}
{Phys. Lett. B \textbf{753}, 18 (2016)}.

\bibitem{Dokshitzer:1977sg}
Y.~L.~Dokshitzer,
Sov. Phys. JETP \textbf{46}, 641 (1977).

\bibitem{Gribov:1972ri}
V.~N.~Gribov and L.~N.~Lipatov,
Sov. J. Nucl. Phys. \textbf{15}, 438 (1972).

\bibitem{Altarelli:1977zs}
G.~Altarelli and G.~Parisi,
\href{https://doi.org/10.1016/0550-3213(77)90384-4}
{Nucl. Phys. \textbf{B126}, 298 (1977)}.

\bibitem{Kang:2017mda}
Z.~B.~Kang, F.~Ringer and W.~J.~Waalewijn,
\href{https://doi.org/10.1007/JHEP07(2017)064}
{J. High Energy Phys. 07 (2017) 064}.

\bibitem{Werner:private}
W. Vogelsang, 
private communication (2021).

\bibitem{Blazey:2000qt}
G.~C.~Blazey \textit{et al.},
\href{https://arxiv.org/abs/hep-ex/0005012}
{arXiv:hep-ex/0005012}.

\bibitem{Ellis:1993tq}
S.~D.~Ellis and D.~E.~Soper,
\href{https://doi.org/10.1103/PhysRevD.48.3160}
{Phys. Rev. D \textbf{48}, 3160 (1993)}.

\bibitem{Cacciari:2008gp}
M.~Cacciari, G.~P.~Salam and G.~Soyez,
\href{https://doi.org/10.1088/1126-6708/2008/04/063}
{J. High Energy Phys. 04 (2008) 063}.

\bibitem{Vogt:2004ns}
A.~Vogt,
\href{https://doi.org/10.1016/j.cpc.2005.03.103}
{Comput. Phys. Commun. \textbf{170}, 65 (2005)}.

\bibitem{Floratos:1981hs}
E.~G.~Floratos, C.~Kounnas and R.~Lacaze,
\href{https://doi.org/10.1016/0550-3213(81)90434-X}
{Nucl. Phys. \textbf{B192}, 417 (1981)}.

\bibitem{Stratmann:2001pb}
M.~Stratmann and W.~Vogelsang,
\href{https://doi.org/10.1103/PhysRevD.64.114007}
{Phys. Rev. D \textbf{64}, 114007 (2001)}.

\bibitem{Cocuzza:2021cbi}
C.~Cocuzza, W.~Melnitchouk, A.~Metz and N.~Sato,
\href{http://doi.org/10.1103/PhysRevD.104.074031}
{Phys. Rev. D \textbf{104}, 074031 (2021)}.

\bibitem{Cocuzza:2021rfn}
C.~Cocuzza, C.~E.~Keppel, H.~Liu, W.~Melnitchouk, A.~Metz, N.~Sato and A.W. Thomas,
\href{http://doi.org/10.1103/PhysRevLett.127.242001}
{Phys. Rev. Lett. \textbf{127}, 242001 (2021)}.


\bibitem{Benvenuti:1989rh}
A. C. Benvenuti {\it et al.},
\href{https://doi.org/10.1016/0370-2693(89)91637-7}
{Phys. Lett. B {\bf 223}, 485 (1989)}.

\bibitem{Whitlow:1991uw}
L.~W.~Whitlow \textit{et al.},
\href{https://doi.org/10.1016/0370-2693(92)90672-Q}
{Phys. Lett. B \textbf{282}, 475 (1992)}.

\bibitem{Arneodo:1996qe}
M.~Arneodo \textit{et al.},
\href{https://doi.org/10.1016/S0550-3213(96)00538-X}
{Nucl. Phys. \textbf{B483}, 3 (1997)}.

\bibitem{Arneodo:1996kd}
M.~Arneodo \textit{et al.},
\href{https://doi.org/10.1016/S0550-3213(96)00673-6}
{Nucl. Phys. \textbf{B487}, 3 (1997)}

\bibitem{Abramowicz:2015mha}
H.~Abramowicz \textit{et al.},
\href{https://doi.org/10.1140/epjc/s10052-015-3710-4}
{Eur. Phys. J. C \textbf{75}, 580 (2015)}.

\bibitem{Ashman:1989ig}
J.~Ashman \textit{et al.},
\href{https://doi.org/10.1016/0550-3213(89)90089-8}
{Nucl. Phys. \textbf{B328}, 1 (1989)}.

\bibitem{Adeva:1998vv}
B.~Adeva \textit{et al.},
\href{https://doi.org/10.1103/PhysRevD.58.112001}
{Phys. Rev. D \textbf{58}, 112001 (1998)}.

\bibitem{Adeva:1999pa}
B.~Adeva \textit{et al.},
\href{https://doi.org/10.1103/PhysRevD.60.072004}
{Phys. Rev. D \textbf{60}, 072004 (1999)}
[Erratum: \href{https://doi.org/10.1103/PhysRevD.62.079902}
{Phys. Rev. D \textbf{62}, 079902 (2000)}].

\bibitem{Alekseev:2010hc}
M.~G.~Alekseev \textit{et al.},
\href{https://doi.org/10.1016/j.physletb.2010.05.069}
{Phys. Lett. B \textbf{690}, 466 (2010)}.

\bibitem{Alexakhin:2006oza}
V.~Y.~Alexakhin \textit{et al.},
\href{https://doi.org/10.1016/j.physletb.2006.12.076}
{Phys. Lett. B \textbf{647}, 8 (2007)}.




\bibitem{Baum:1983ha}
G.~Baum \textit{et al.}
\href{https://doi.org/10.1103/PhysRevLett.51.1135}
{Phys. Rev. Lett. \textbf{51}, 1135 (1983)}.

\bibitem{Anthony:1996mw}
P.~L.~Anthony \textit{et al.},
\href{https://doi.org/10.1103/PhysRevD.54.6620}
{Phys. Rev. D \textbf{54}, 6620 (1996)}.

\bibitem{Abe:1998wq}
K.~Abe \textit{et al.},
\href{https://doi.org/10.1103/PhysRevD.58.112003}
{Phys. Rev. D \textbf{58}, 112003 (1998)}.

\bibitem{Abe:1997cx}
K.~Abe \textit{et al.},
\href{https://doi.org/10.1103/PhysRevLett.79.26}
{Phys. Rev. Lett. \textbf{79}, 26 (1997)}.


\bibitem{Anthony:2000fn}
P.~L.~Anthony \textit{et al.},
\href{https://doi.org/10.1016/S0370-2693(00)01014-5}
{Phys. Lett. B \textbf{493}, 19 (2000)}.

\bibitem{Anthony:1999rm}
P.~L.~Anthony \textit{et al.},
\href{https://doi.org/10.1016/S0370-2693(99)00940-5}
{Phys. Lett. B \textbf{463}, 339 (1999)}.



\bibitem{Ackerstaff:1997ws}
K.~Ackerstaff \textit{et al.},
\href{https://doi.org/10.1016/S0370-2693(97)00611-4}
{Phys. Lett. B \textbf{404}, 383 (1997)}.

\bibitem{Airapetian:2007mh}
A.~Airapetian \textit{et al.},
\href{https://doi.org/10.1103/PhysRevD.75.012007}
{Phys. Rev. D \textbf{75}, 012007 (2007)}.

\bibitem{Sato:2016tuz}
N.~Sato, W.~Melnitchouk, S.~E.~Kuhn, J.~J.~Ethier, and A.~Accardi,
\href{http://doi.org/10.1103/PhysRevD.93.074005}
{Phys. Rev. D \textbf{93}, 074005 (2016)}.

\bibitem{Hawker:1998ty}
E.~A.~Hawker \textit{et al.},
\href{https://doi.org/10.1103/PhysRevLett.80.3715}
{Phys. Rev. Lett. \textbf{80}, 3715 (1998)}.

\bibitem{Alekhin:2006zm}
S.~Alekhin, K.~Melnikov and F.~Petriello,
\href{https://doi.org/10.1103/PhysRevD.74.054033}
{Phys. Rev. D \textbf{74}, 054033 (2006)}.

\bibitem{Abazov:2008ae}
V.~M.~Abazov \textit{et al.},
\href{https://doi.org/10.1103/PhysRevLett.101.062001}
{Phys. Rev. Lett. \textbf{101}, 062001 (2008)}.

\bibitem{Abulencia:2007ez}
A.~Abulencia \textit{et al.},
\href{https://doi.org/10.1103/PhysRevD.75.092006}
{Phys. Rev. D \textbf{75}, 092006 (2007)}
[Erratum: \href{https://doi.org/10.1103/PhysRevD.75.119901}
{Phys. Rev. D \textbf{75}, 119901 (2007)}].

\bibitem{zhou:thesis}
Y.~Zhou,
\href{https://doi.org/10.21220/9K5N-4T96}
{Dissertations, Theses, and Masters Projects, William \& Mary, Paper 1627407586}.

\bibitem{AbdulKhalek:2021gbh}
R.~Abdul Khalek \textit{et al.},
\href{https://arxiv.org/abs/2103.05419}
{arXiv:2103.05419 [physics.ins-det]}.

\bibitem{Jimenez-Delgado:2013boa}
P.~Jimenez-Delgado, A.~Accardi and W.~Melnitchouk,
\href{https://doi.org/10.1103/PhysRevD.89.034025}
{Phys. Rev. D \textbf{89}, 034025 (2014)}.

\bibitem{Sato:2016wqj}
N.~Sato, J.~J.~Ethier, W.~Melnitchouk, M.~Hirai, S.~Kumano, and A.~Accardi,
\href{http://doi.org/10.1103/PhysRevD.94.114004}
{Phys. Rev. D \textbf{94}, 114004 (2016)}.

\bibitem{NNPDF:2017mvq}
R.~D.~Ball \textit{et al.},
\href{https://doi.org/10.1140/epjc/s10052-017-5199-5}
{Eur. Phys. J. C \textbf{77}, 663 (2017)}.

\bibitem{Cooper-Sarkar:2018ufj}
A.~M.~Cooper-Sarkar and K.~Wichmann,
\href{https://doi.org/10.1103/PhysRevD.98.014027}
{Phys. Rev. D \textbf{98}, 014027 (2018)}.

\bibitem{Harland-Lang:2014zoa}
L.~A.~Harland-Lang, A.~D.~Martin, P.~Motylinski and R.~S.~Thorne,
\href{https://doi.org/10.1140/epjc/s10052-015-3397-6}
{Eur. Phys. J. C \textbf{75}, 204 (2015)}.

\bibitem{Egan75}
J.~P.~Egan,
{\it Signal detection theory and ROC analysis},
Series in Cognition and Perception
(Academic Press, New York, 1975).

\bibitem{Fawcett06}
T.~Fawcett,
\href{https://doi.org/10.1016/j.patrec.2005.10.010}
{Pattern Recogn. Lett. {\bf 27}, 861 (2006)}.

\bibitem{Whitehill22}
R.~M.~Whitehill, Y.~Zhou, N.~Sato and W.~Melnitchouk,
in preparation (2022).

\bibitem{COMPASS:2012mpe}
C.~Adolph \textit{et al.} [COMPASS],
\href{https://doi.org/10.1103/PhysRevD.87.052018}
{Phys. Rev. D \textbf{87}, 052018 (2013).}

\bibitem{Anderle:2021hpa}
D.~P.~Anderle, X.~Dong, F.~Hekhorn, M.~Kelsey, S.~Radhakrishnan, E.~Sichtermann, L.~Xia, H.~Xing, F.~Yuan and Y.~Zhao,
\href{https://doi.org/10.1103/PhysRevD.104.114039}
{Phys. Rev. D \textbf{104}, 114039 (2021).}


\end{thebibliography}
\end{document}